\documentclass[aip,graphicx]{revtex4-1}

\usepackage[english]{babel}
\usepackage[utf8]{inputenc}
\usepackage{graphicx}
\usepackage{dcolumn}
\usepackage{bm}
\usepackage{amsmath}
\usepackage{amssymb}
\usepackage{datetime}
\usepackage{hyperref}
\usepackage{bigints}
\usepackage{siunitx}
\usepackage[caption=false]{subfig}

\begin{document}

\title{On the mechanism of ionization oscillations in Hall thrusters}
\author{\firstname{O.}~\surname{Chapurin}}
\email{alex.chapurin@usask.ca}
\affiliation{Department of Physics and Engineering Physics, University of Saskatchewan, Saskatoon SK S7N 5E2, Canada}
\author{\firstname{A. I.}~\surname{Smolyakov}}
\affiliation{Department of Physics and Engineering Physics, University of Saskatchewan, Saskatoon SK S7N 5E2, Canada}
 
\author{\firstname{G.}~\surname{Hagelaar}}
\affiliation{LAPLACE, Université de Toulouse, CNRS, INPT, UPS, 118 Route de Narbonne, 31062 Toulouse, France}
\author{\firstname{Y.}~\surname{Raitses}}
\affiliation{Princeton Plasma Physics Laboratory, Princeton, New Jersey 08540, USA}

\begin{abstract}
Low frequency ionization oscillations involving plasma and neutral density (breathing modes) are the most violent perturbations in Hall thrusters for electric propulsion. Because of its simplicity, the zero-dimensional (0-D)  predator-prey model of two nonlinearly coupled ordinary differential equations for plasma and neutral density has been often used for the characterization of such oscillations and scaling estimates. We investigate the properties of its continuum analog, the one-dimensional (1-D) system of two nonlinearly coupled equations in partial derivatives  (PDE) for plasma and neutral density. This is a more general model, of which the standard 0-D predator-prey model is a special limit case.  We show that the 1-D model is stable and does not show any oscillations for the boundary conditions relevant to Hall thruster and the uniform ion velocity. We then propose a reduced 1-D model based on two coupled PDE for plasma and neutral densities that is unstable and exhibit oscillations if the ion velocity profile with the near the anode back-flow (toward the anode) region is used.  Comparisons of the reduced model with the predictions of the full model that takes into account the self-consistent plasma response show that the main properties of the breathing mode are well captured. In particular, it is shown that the frequency of the breathing mode oscillations is weakly dependent on the final ion velocity but shows a strong correlation with the width of the ion back-flow region.   
\end{abstract}

\maketitle


\section{Introduction}

Hall thrusters are electric propulsion plasma devices with cross-field $E\times B$ configuration, where electrons are effectively trapped by the magnetic field, and heavy ions are accelerated by the electric field generating thrust. Despite the long history and a large number of
successful missions, as well as the relatively simple design of these devices,
many physical phenomena are not well understood. Large amplitude discharge
current oscillations in the axial direction of Hall thrusters, or the
so-called breathing modes, are among the most strong perturbations that
affect the operation, e.g.\ with current oscillations reaching 100\% and even extinguishing the discharge\cite{zhurin1993dynamic}. However note that some modifications of Hall thrusters exhibit no or very little low-frequency oscillations\cite{raitses2001parametric,liqiu2012study}. The physics of the breathing mode has been studied in
many papers but the full understanding is far from complete.  Simulations of low-frequency axial dynamics of Hall
thrusters were done earlier with various approximations, including fluid\cite%
{morozov2000fundamentals, hagelaar2002two, barral2009low}, and hybrid\cite{morozov2000fundamentals, boeuf1998low, fife1998hybrid, chable2005numerical, hara2012one}
(fluid electrons and kinetic ions and atoms) models.  Ionization is
one of the key physics elements of the breathing mode oscillations. In addition to the basic coupling of plasma density and neutral gas due to the ionization, many models include
self-consistent electric field calculations along the channel, and some models include electron pressure evolution and detailed electron energy balance.  

At the same time, a simple  zero-dimensional (0-D) model was proposed \cite{fife1997numerical,
barral2008origin} to explain the mechanism of breathing mode, in which
neutral and ion (plasma) densities are coupled in the form of predator-prey
model (Lotka-Volterra equations).
The predator-prey model is a system of two nonlinearly coupled ordinary
differential equations that describes periodic oscillations in a number of
physics and biology systems. This model is claimed to predict the frequencies close to the observed in experiments. It was, however, noted that for the constant ion and neutral velocities,  the model is stable and does not predict any 
instability  \cite{hara2014perturbation}  so the conditions for the excitation
of the ionization oscillations are not clear. Additional effects and modifications  \cite{hara2014perturbation,Wang2010CPP}
were proposed for the predator-prey model to make it unstable but it remains unclear  to what degree
the 0-D models are capable to predict the breathing mode oscillations (discussed in section~\ref{sec_0dpp}).

Nevertheless, simple reduced models are of considerable interest. First, such models are useful as scaling tools, e.g.\ for the design of power supply sources. Another motivation stems from the considerable complexity of the full models. In the full models with many parameters that are unknown experimentally, it is rather difficult to predict the conditions for the instability and therefore difficult to translate the predictions of the theoretical models into practical recommendations.  The search for physical mechanism (or mechanisms) and critical controlling parameters continues \cite{dale2019frequency}. Reduced models focusing on selected specific phenomena may allow easier experimental validation and eventually lead to a better understanding of the physics of the instability, its conditions, and the development of the full predictive model(s) \cite{dale2019two1,dale2019two2}.

In this work, we analyze the properties of the continuum one-dimensional (1-D) model, based on two coupled partial differential equations (PDE) for the evolution of neutral and plasma densities. The 0-D predator-prey is a reduction of the 1-D model. If anything, the more complete 1-D model should exhibit the same physics (and e.g. predict similar characteristics of the oscillations)   if the reduction from the 1-D to the 0-D model is appropriate for the relevant  physics, and the oscillations in the 0-D model are not the artefact introduced by the simplifications. 
We show, however, that the 1-D model is not only stable with boundary conditions of a typical Hall thruster and the standard assumptions of the {\it uniform velocity} (as it is used in 0-D model) but also does not show the oscillations observed in the 0-D model. On the contrary, any oscillations, even introduced externally,  are damped in time and space.  We then propose 
a modified 1D model with a non-uniform  ion velocity profile and  with a back-flow  
(toward the anode) the region near the anode. We show that such a model is unstable and exhibit
ionization oscillations with characteristics roughly consistent with the predictions of the full fluid
model\cite{smolyakov2019theory}. Rather good agreement of the results of the reduced (simplified) and full models suggests that the reduced model captures well essential physics relevant to the breathing mode and may point to the critical conditions for the instability. 

The paper is organized as follows. In section~\ref{sec_0dpp}, various modifications of the 0-D model are discussed. In section \ref{sec_1dpp}, the continuum 1-D model is introduced. The stationary solutions are obtained here, their stability and response to external perturbations are discussed for the uniform profile of the ion velocity. Section \ref{sec_1doscillations} presents a reduced 1-D continuum model with the ion  back-flow region near the anode and shows that this model exhibit unstable nonlinear  oscillations. In section  \ref{sec_comparison}, the predictions  of the reduced model are compared against the full self-consistent simulations.


\section{Zero-dimensional (0-D) predator-prey models} \label{sec_0dpp}

In this section we discuss the various modifications and properties of the zero-dimensional (0-D) predator-prey models for the ionization oscillations. 
Such a simplest model  includes basic time
and space dynamics of plasma and neutrals and is  based on the  continuity equations for the ion
density $n_{i}$ (assuming full plasma quasineutrality), and neutral atom
density $n_{a}$: 
\begin{eqnarray}
&&\frac{\partial n_{a}}{\partial t}+v_{a}\frac{\partial n_{a}}{\partial x}
=-\beta n_{a}n_i,  \label{na_cont} \\
&&\frac{\partial n_{i}}{\partial t}+\frac{\partial }{\partial x}\left(
n_{i}v_{i}\right) =\beta n_{a}n_{i},  \label{ni_cont}
\end{eqnarray}%
where $v_{i},v_{a}$ are ion and atom flow velocities, respectively, and $%
\beta $ is the ionization rate coefficient. We assume that $v_{a}$ is a
constant, but $v_{i}$ and $\beta $, in general, could be non-uniform in
space. A basic assumption leading to a simple
zero-dimensional predator-prey model \cite{fife1997numerical} is that the
ionization zone and acceleration zone is replaced by the transition layer of
the width $L$. The integration of Eqs.~(\ref{na_cont})-(\ref{ni_cont}) over
this region results in the system of ordinary differential equations (ODE):
\begin{eqnarray}
&&\frac{\partial }{\partial t}\left\langle n_{a}\right\rangle +\frac{1}{L}%
\left. \left( n_{a}v_{a}\right) \right\vert _{0}^{L}=-\beta \left\langle
n_{a}n_{i}\right\rangle, \\
&&\frac{\partial }{\partial t}\left\langle n_{i}\right\rangle +\frac{1}{L}%
\left. \left( n_{i}v_{i}\right) \right\vert _{0}^{L}=\beta \left\langle
n_{a}n_{i}\right\rangle,
\end{eqnarray}%
where averages over the transition layer are defined as $\left\langle \left(
...\right) \right\rangle =L^{-1}\int_{0}^{L}\left( ...\right) dx$ $\ $\, and
$\beta $ is assumed constant here.  A set of approximations are made  to obtain   the
original predator-prey model\cite{fife1997numerical}. \ First, the ion and
neutral fluxes at the boundaries of the transition layer are assumed in the
form: $\left. \left( n_{i}v_{i}\right) \right\vert _{0}=0$, $\left. \left(
n_{i}v_{i}\right) \right\vert _{L}=nv_{i}$ and $\left. \left(
n_{a}v_{a}\right) \right\vert _{0}=n_{a}v_{a}$ and $\left. \left(
n_{a}v_{a}\right) \right\vert _{L}=0$, where $v_{i}$ is the final ion
velocity and $n(t)$ is plasma density at the exit of the
transition layer, at $x=L$, $n(t) \equiv n_{i}(t,L)$, 
and $N(t) $ is neutral density at $x=0$, $N(
t) \equiv n_{a}(t,0)$. Thus, these boundary conditions
imply full ionization, $n_{a}(L) =0$,  zero flux of the ions from the left boundary,$\left. \left( n_{i}v_{i}\right) \right\vert _{0}=0$, and full acceleration the transition layer, $\left. \left( n_{i}v_{i}\right) \right\vert _{L}=nv_{i}$. Then, the boundary values
for the plasma and neutral densities are used to approximate the averages
over the transition layer: $\left\langle n_{a}n_{i}\right\rangle \simeq
nN$, $\left\langle n_{a}\right\rangle \simeq N$, and $\left\langle
n_{i}\right\rangle \simeq n$.  These steps lead to the 0-D ODE system \cite%
{fife1997numerical}
\begin{eqnarray}
&&\frac{d N}{d t}-\frac{1}{L}Nv_{a}=-\beta Nn,
\label{p1} \\
&&\frac{d n}{d t}+\frac{1}{L}nv_{i}=\beta Nn.  \label{p2}
\end{eqnarray}

These equations have the equilibrium solution with $n_{eq}=v_{a}/L\beta $
and $N_{eq}=v_{i}/L\beta $. Considering the perturbations $\left(
\widetilde{n}(t), \widetilde{N}(t) \right) \sim
\exp \left( -i\omega t\right) $ near the equilibrium, $n=$ $n_{eq}+%
\widetilde{n}(t)$, $N=$ $N_{eq}+\widetilde{N}(
t) $ one obtains stable oscillations with 
\begin{equation}
  \omega ^{2}=\beta
^{2}n_{eq}N_{eq}=v_{i}v_{a}/L^{2}
\label{pp}
\end{equation}

It was later noted \cite%
{barral2008origin} that oscillations (time dependence) in $N$ are inconsistent with the
constant mass rate boundary condition at $x=0$,  $\left. \left(
n_{a}v_{a}\right) \right\vert _{0}=Nv_{a}=\dot{M}%
/Am_{a}=const,$ which means that the $N$ should be constant for the constant neutral flow velocity;  $A$ is the cross-section area. 
A  modification of the 0-D model was suggested in Ref.~\onlinecite{hara2014perturbation} where the value of neutral density at the left
(entrance) side was fixed constant, $\left. \left( n_{a}v_{a}\right) \right\vert
_{0}=N_{0}v_{a}$, and the  value at the exit was assumed time dependent, $%
\left. \left( n_{a}v_{a}\right) \right\vert _{L}=N\left( t\right) v_{a},$
so the neutral balance equation takes the form
\begin{equation}
\frac{d N}{d t}+\frac{1}{L}\left( N-N_{0}\right)
v_{a}=-\beta Nn.  \label{m1}
\end{equation}%
The plasma balance equation was also modified with an additional term due to
the sheath losses at the lateral walls
\begin{equation}
\frac{d n}{d t}+\frac{1}{L}nv_{i}+\frac{1}{d}nc_{s}=\beta
Nn,  \label{m2}
\end{equation}%
where sheath losses are estimated based on the Bohm condition for the
ion velocity, where $d$ is the radial channel width and $c_s=\sqrt{T_e/m_i}$. With these modifications, the equilibrium value of the neutral
density in equations~(\ref{m1}-\ref{m2}) is only corrected by  the  sheath
losses,
\begin{equation}
N_{eq}=\frac{1}{\beta }\left( \frac{v_{i}}{L}+\frac{c_{s}}{d}\right),
\end{equation}%
while the equilibrium value of plasma density can be very different from
that in equation (\ref{p1}) and takes the form%
\begin{equation}
n_{eq}=\frac{v_{a}}{\beta L}\left( \frac{N_{0}}{N_{eq}}-1\right).
\end{equation}%
Note that in this model, the value of the neutral density at $x=0$ should be sufficiently large $N_{0}>N_{eq}=\left(
v_{i}/L+c_{s}/d\right) /\beta $. Considering perturbations near  $N_{eq}$
and $n_{eq}$ one obtains damped oscillations \cite{hara2014perturbation}%
\begin{equation}
\omega =-\frac{i}{2}\frac{v_{a}}{L}\frac{N_{0}}{N_{eq}}\pm \sqrt{\beta
^{2}n_{eq}N_{eq}-\frac{1}{4}\left( \frac{v_{a}}{L}\right) ^{2}\left( \frac{%
N_{0}}{N_{eq}}\right) ^{2}}.
\end{equation}

The essential problem of the predator-prey model is that it does not
predict instability so no condition for the oscillations can be determined.
It was argued that the electron dynamics should be important and more complex models were proposed  such as using
the electron (instead of ion) continuity equation with the drift-diffusion
approximation for the electron velocity \cite%
{yamamoto2002discharge,yamamoto2005discharge,yamamoto2005suppression}, ion flows \cite{Wang2010CPP}, electron energy evolution \cite{hara2014perturbation,Hara2017DirectKS}, and
two-zone model \cite{dale2019two1, dale2019two2}. Ref. \onlinecite{WeiChinPhysB2015} provides an overview of various models and involved mechanisms.


\section{Continuum (1-D) predator-prey model} \label{sec_1dpp}

\begin{figure}[htbp]
\centering
\subfloat[$n_0 / n_{a0} = 0.05, \ l_0 = \SI{1.31}{cm}$]{\includegraphics[width=0.5\textwidth]{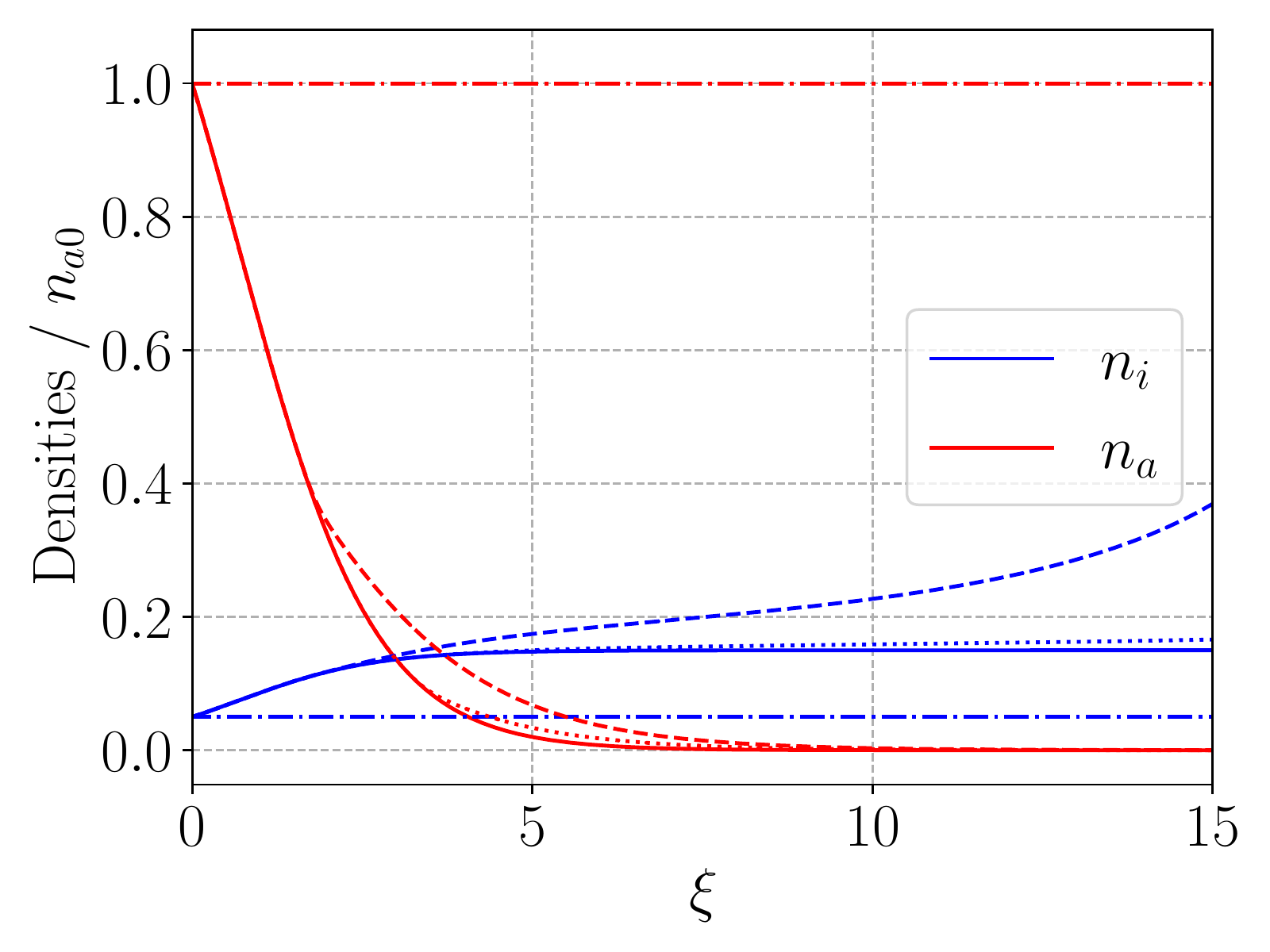}\label{stnr_sols1}}
\subfloat[$n_0 / n_{a0} = 0.0001, \ l_0 = \SI{ 1.96}{cm}$]{\includegraphics[width=0.5\textwidth]{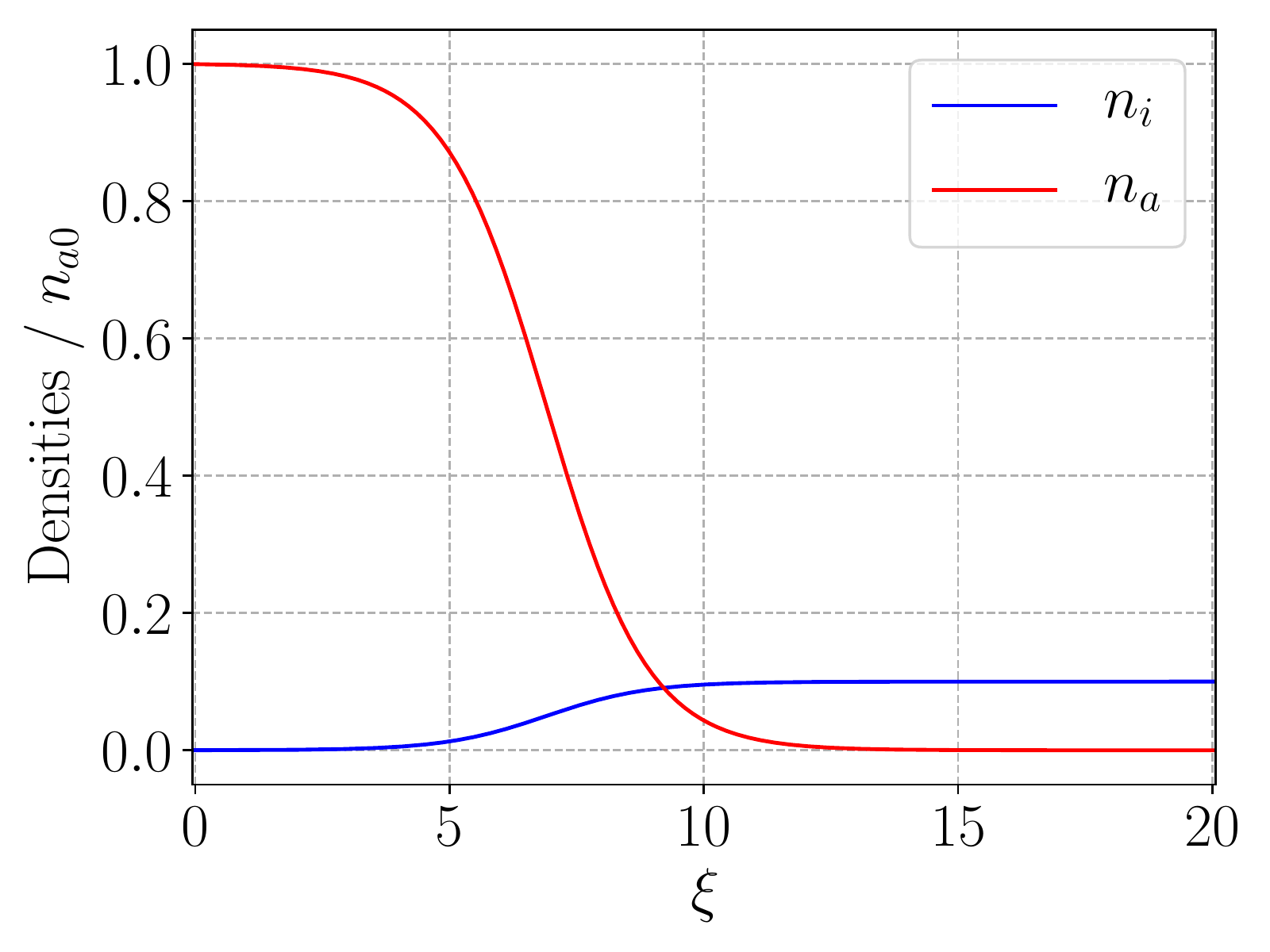}\label{stnr_sols2}}
\caption{Stationary solutions given by Eqs.~(\ref{na_stnry}) and (\ref{ni_stnry}) are shown by solid lines, with higher (a) and lower (b) ion density at the left boundary $n_0$. The time dependent solutions of Eqs.~(\ref{na_cont}), and (\ref{ni_cont}), converging to the stationary solution are shown for the case (a) at  three consecutive time steps $t_1 < t_2 < t_3$, dot dashed (initial condition), dashed, and dotted line, respectively. }
\label{stnr_sols}
\end{figure}

Ideally, the reduction from the continuous 1-D model to the 0-D models as was discussed in the previous section should preserve the essential properties  and features of the more complete model  and do not introduce  any fundamental changes to the 1-D model. Therefore,  is of interest to study if the 1-D model has the properties predicted by the 0-D models.  Here we discuss properties of the stationary and time-dependent solutions of
the one-dimensional (continuous) model consisting of the continuity
equations~(\ref{na_cont}) and (\ref{ni_cont}) for atoms and ions with constant
flow velocities $v_{a}$ and $v_{i}$, and the ionization rate $\beta $. These are the same assumptions with which the original predator-prey model was derived.  The
stationary problem is formulated as two coupled equations:
\begin{eqnarray}
&&v_{a}\frac{\partial n_{a}}{\partial x}=-\beta n_{a}n_i,  \label{na_stnry} \\
&&v_{i}\frac{\partial n_{i}}{\partial x}=\beta n_{a}n_{i}.  \label{ni_stnry}
\end{eqnarray}%
The exact solution to this system is found in the form:
\begin{eqnarray}
&&n_{a,st}(\xi )=n_{a0}\left( 1+\frac{n_{0}v_{i}}{n_{a0}v_{a}}\right) \dfrac{%
1}{\left( n_{0}v_{i}\right) /\left( n_{a0}v_{a}\right) \exp {\left( \xi
\right) }+1},  \label{na_st_sol} \\
&&n_{i,st}(\xi )=n_{0}\left( 1+\frac{n_{a0}v_{a}}{n_{0}v_{i}}\right) \dfrac{%
\exp {\left( \xi \right) }}{\exp {\left( \xi \right) }+\left(
n_{a0}v_{a}\right) /\left( n_{0}v_{i}\right) },  \label{ni_st_sol}
\end{eqnarray}%
where $n_{0},n_{a0}$ are values of ion density and atom density,
respectively, at the left boundary, and normalized length is $\xi =x/l_{0}$,
with $l_{0}=v_{i}v_{a}/\beta \left( n_{0}v_{i}+n_{a0}v_{a}\right) $. It can
be noted that the solutions of the stationary problem depend only on two
parameters, $n_{0},n_{a0}$. Typical stationary solutions for $v_i/v_a = 10$ are depicted in
Figs.~\ref{stnr_sols1},\ref{stnr_sols2} by solid lines. Note that the position of the crossing point, where the ion and neutral densities are equal, and the localization of the ionization source $\beta nn_{a}$ depend on the value $n_{0}$: they move
to the right with decreasing  $n_{0}$, and goes to infinity for $
n_{0} \to 0$. 

Numerical studies of the time-dependent equations~(\ref{na_cont}) and (\ref{ni_cont}) were performed with fixed boundary values at the left $n_{0},n_{a0}$ and free boundary conditions (spatial second derivative is zero) on the right end. This analysis shows that all perturbations are damped and converge to the stationary solutions given by equations~(\ref{na_st_sol}) and~(\ref{ni_st_sol}), as shown in Fig.~\ref{stnr_sols1} at different times. Thus, the 1-D
predator-prey model with a constant spatial profile of the ion velocity does not exhibit oscillatory or unstable behaviour, unlike zero-dimensional predator-prey models.
The 1-D continuum model presented here in some sense is similar to the asymptotic low-frequency model derived in the limit $n/N_a \ll 1$ \cite{barral2008origin,barral2009low}, which also predicts the neutral oscillations with the frequencies  of the order of given by equation (\ref{pp}). Contrary to \cite{barral2008origin,barral2009low}, our model does not show neutrally stable oscillations but only damped modes. 

One can envisage that the boundary value of plasma density could be  perturbed, e.g.\ by some sheath instability. Such perturbations
may also arise in the experiments with segmented anode \cite{RaitsesJAP2006}, external modulations of the anode voltage \cite{Romadanov2018PSST}, and two-stage thruster configurations \cite{FischJAP2001}.  Therefore, it is  interesting to investigate   how this nonlinear system responds to external  perturbations of the ion density and whether such external perturbations may grow in amplitude while propagating from the left boundary. We will impose harmonic external modulations  of the ion density at the left boundary in the form  $n_i(0)=n_{i0}\left( 1+r\sin{2\pi f_{ext}t}\right)$, where $f_{ext}$ is the modulation frequency, with $r=0.15$. We consider the  1-D continuum model, Eqs.~(\ref{na_cont},\ref{ni_cont})
with $v_i = 10 v_a$, $n_{i0}/n_{a0} = 0.05$, $l_0 = \SI{1.31}{cm}$, Fig.~\ref{stnr_sols1}, with the simulation length $15 l_0$. Since velocities of
neutrals and ions are different, there are two natural modes that can be excited during the external modulations, with the wavelengths  given with $\lambda_a = v_a/f_{ext}$ and $\lambda_i = v_i/f_{ext}$. We consider four values of the driving frequencies  $f_{ext} l_0/v_{a} = (0.1, 0.5, 1, 10)$. For these values, the corresponding natural wavelengths of the neutral and ion characteristic wavelengths are $\lambda_a = (10, 2, 1, 0.1) l_0$ and  $\lambda_i = (100, 20, 10, 1) l_0$, respectively. 

\begin{figure}[!htbp]
\centering
\subfloat[$f_{ext} l_0 / v_a = 0.1$]{\includegraphics[width=0.5\textwidth]{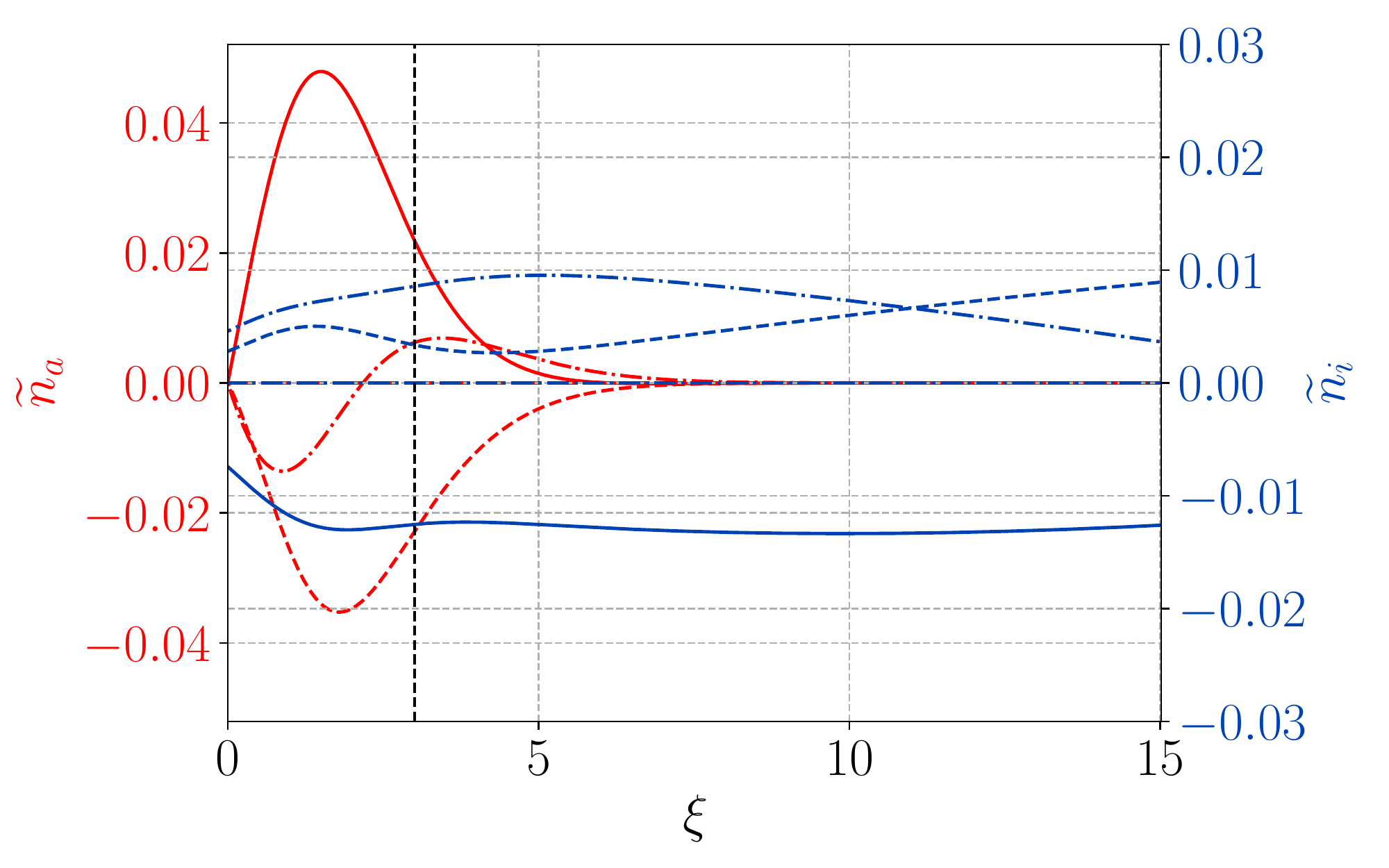}\label{drive_ni_f01}}
\subfloat[$f_{ext} l_0  / v_a = 0.5$]{\includegraphics[width=0.5\textwidth]{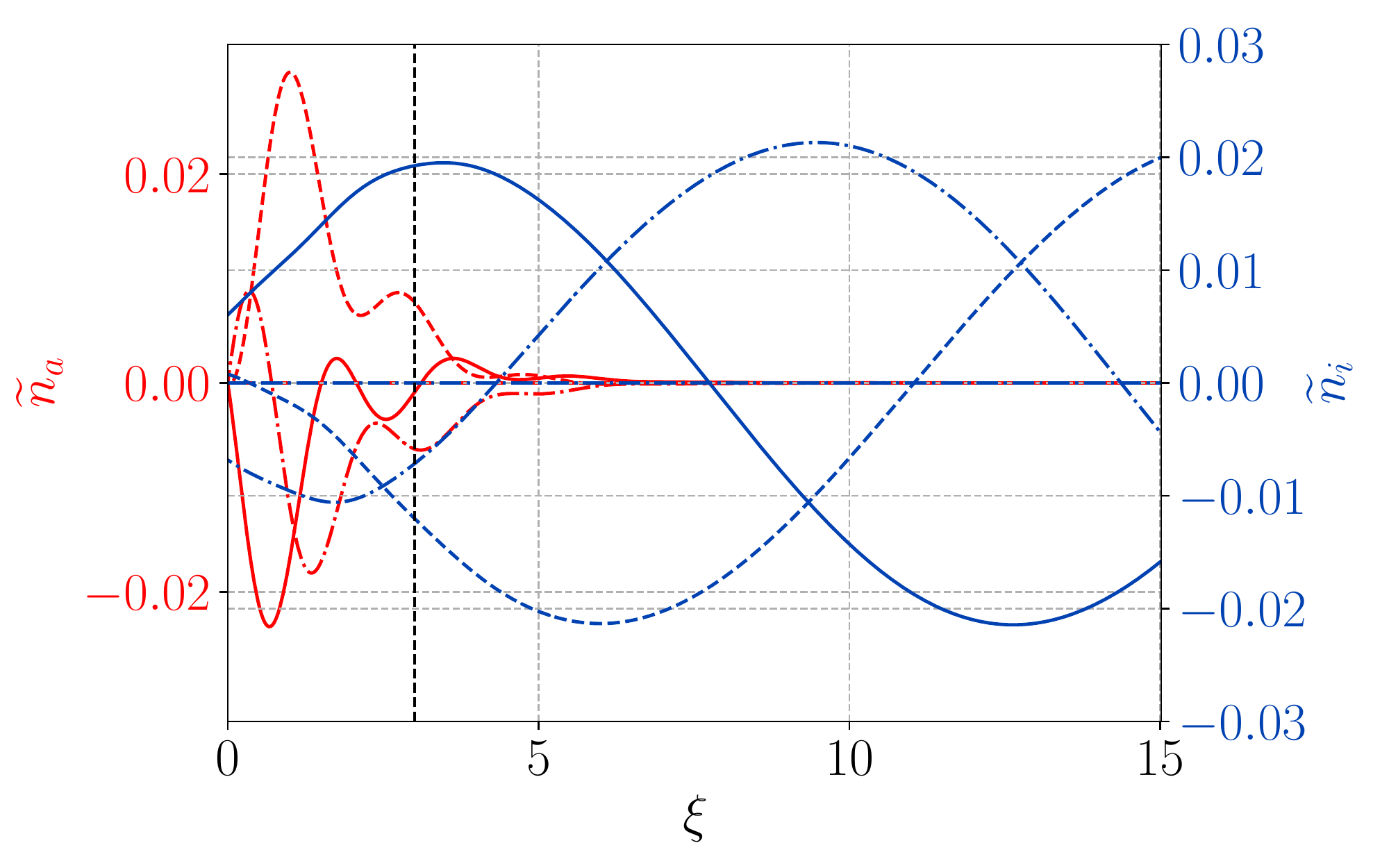}\label{drive_ni_f05}} \\
\subfloat[$f_{ext} l_0  / v_a = 1$]{\includegraphics[width=0.5\textwidth]{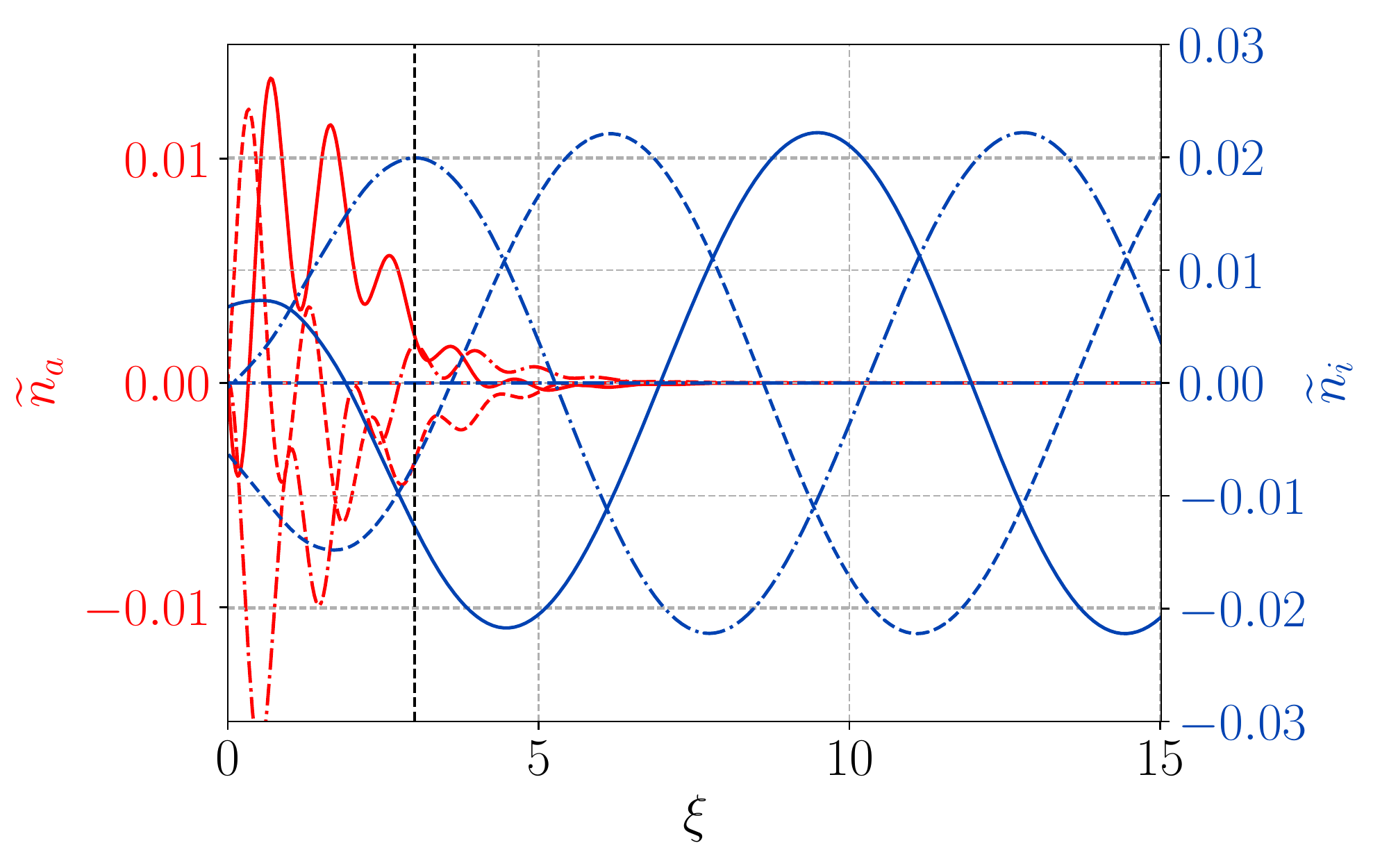}\label{drive_ni_f1}} 
\subfloat[$f_{ext} l_0  / v_a = 10$]{\includegraphics[width=0.5\textwidth]{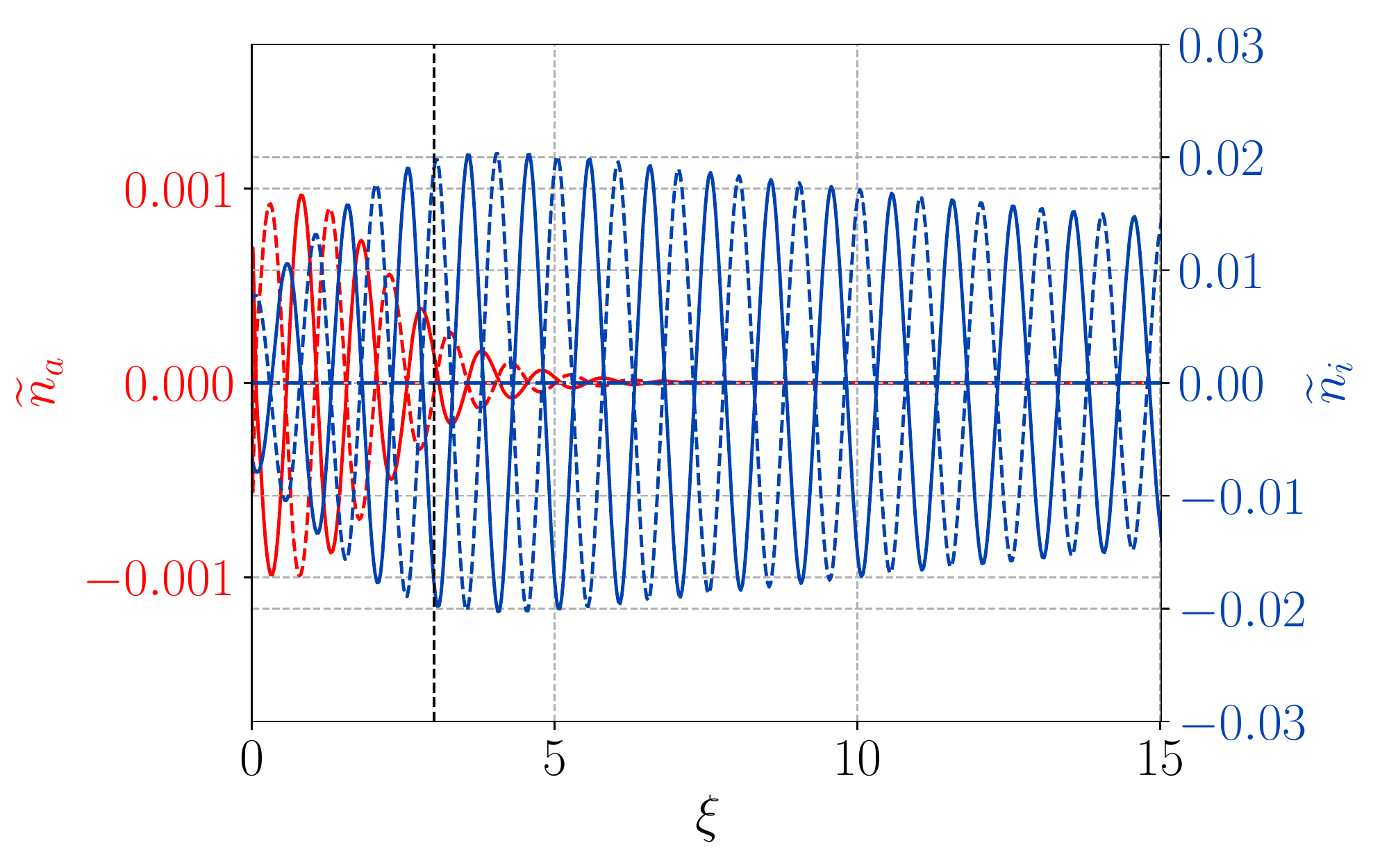}\label{drive_ni_f10}}
\caption{Instantaneous profiles of perturbed ion $\widetilde{n}_i$ (blue) and neutral $\widetilde{n}_a$ (red) densities for the various driving frequencies $f_{ext} l_0  / v_a = (0.1,0.5,1,10)$, shown, respectively, in (a), (b), (c), (d). Three consecutive time snapshots $t_1 < t_2 < t_3$ are shown for each variable, depicted with the dot dashed line, dashed line, and solid line, respectively; they divide one oscillation period in 3 equal intervals. Crossing point (of stationary ion and atom densities profiles) location is shown with the black dashed line.}
\label{drive_ni}
\end{figure}

The resulting response  of the system of Eqs.~(\ref{na_cont},\ref{ni_cont}) under the external modulations of the ion density is presented in Fig.~\ref{drive_ni} for the perturbed variables, defined as $\widetilde{n}=n-n_{st}$, where $n_{st}$ is the corresponding stationary solution. It can be noted that in all cases ion density perturbations grow in absolute values before reaching  the crossing point and leaving  the ionization region, whereas the amplitude of neutral density perturbations decays, which is expected in the ionization region. Note that while the absolute value of the perturbations of the ion density seems increasing, the relative value $\widetilde{n}/n_{st}$  does not grow.  Thus, in the presence of  external ion density perturbations in the near-anode region, the system exhibit externally driven oscillations that are advected
along the channel but show a limited increase of the amplitude. However, the atom response is slightly more complicated and besides the natural mode can exhibit the nonlinear response (on ion natural wavelength). It is also affected by the atom stationary profile (Fig.~\ref{stnr_sols1}), as atom perturbations decay rapidly after the crossing point. It is seen that for the low frequency perturbations the natural atom mode is excited (Fig.~\ref{drive_ni_f01}), while for the higher frequencies the ion characteristic wavelengths become more and more dominant in the atom response (seen gradually through Figs.~\ref{drive_ni_f05}-\ref{drive_ni_f10}). We remind that the amplitude of the ion density perturbations was fixed in these cases. The amplitude of the neutral density perturbations is maximal for the lower frequencies.   We have also investigated the case when the neutral density was modulated at the left boundary. Such a system shows similar behaviour and does not show the unstable modes.


\section{Ionization oscillations in the 1-D continuous model with the ion back-flow in the near anode region} \label{sec_1doscillations}

As it was discussed in the previous section, the reduced fluid model, described with Eqs.~(\ref{na_cont}, \ref{ni_cont}), with spatially uniform profiles of $v_i$, $v_a$, and $\beta$ is stable. Here we will show that the same system with spatial dependence of $v_i$, when $v_i$ is negative near the anode, the so-called ion back-flow, becomes unstable and exhibit self-consistent oscillations. The ion back-flow region naturally forms in the presheath, a quasineutral region near the anode with the negative electric field due to the electron diffusion\cite{CohenZurPoP2002,AhedoPoP2005,dorf2003anode, dorf2005experimental}. The positive density gradient near the anode creates a strong electron flow to the anode. To maintain ambipolarity (the total current remains uniform), the ion back-flow current occurs to compensate the increase of the electron current. The current ambipolarity is required to keep plasma quasineutral. This process is somewhat similar to the formation of the electric field in the presheath region near the plasma boundary where the electric field is induced to accelerate ions toward the boundary (up to the  Bohm velocity) to balance electron and ion losses. 

\begin{figure}[htbp]
\centering
\subfloat[]{\includegraphics[width=0.5\textwidth]{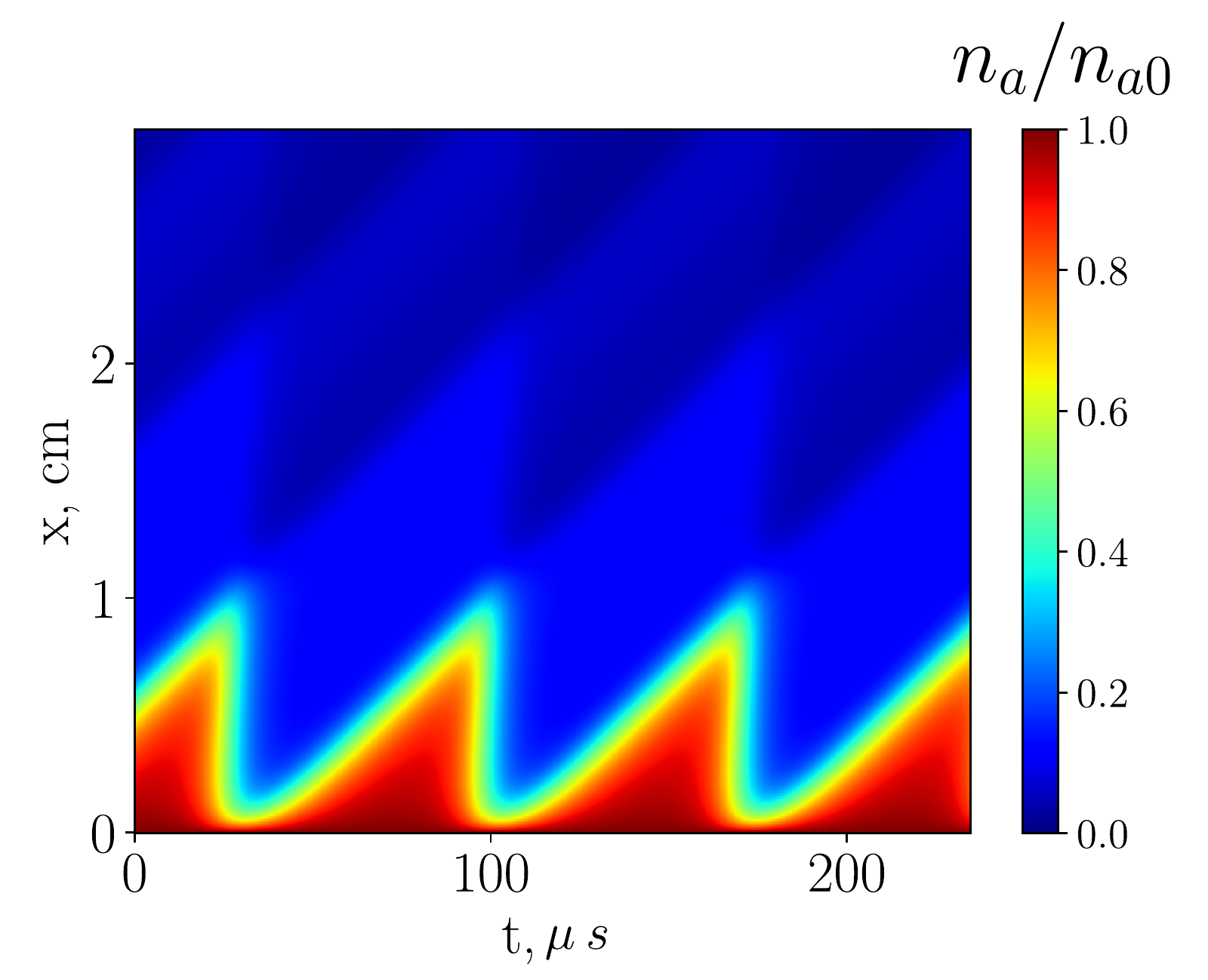}\label{nn_osc}}
\subfloat[]{\includegraphics[width=0.5\textwidth]{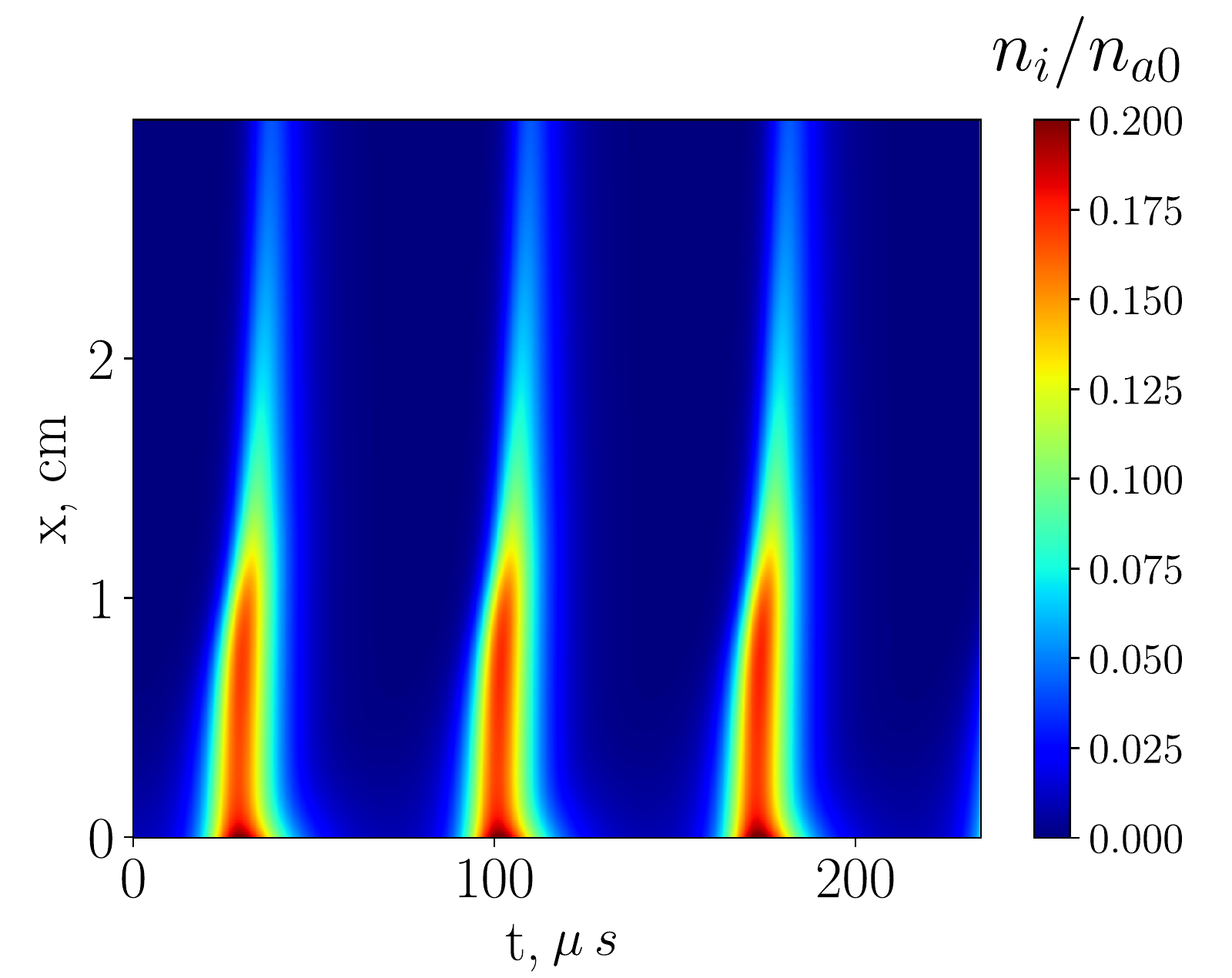}\label{ni_osc}}
\caption{Evolution in space and time of atom (a) and ion (b) density, normalized to $n_{a0}$.}
\label{dens_osc}
\end{figure}

In this configuration the only fixed boundary condition is a value of the neutral density on the left, as their characteristics travel strictly to the right (atom velocity is a positive constant). Ion density boundary condition is free on the left, $\partial^2_x n_i (0) = 0$. First, we consider the simple configuration with the velocity profile $v_i$  as a strictly linear function of the position. The ionization rate $\beta$ is taken uniform. One might argue the constant $\beta$ is an unrealistic assumption for a typical Hall thruster where electron energy near the anode is low due to higher mobility, shaping the $\beta$  profile; we consider these effects in Section~\ref{sec_comparison} in the comparison with the full fluid model. Recombination of plasma at the anode is not included here: the ion flux converted to the neutral flux at the anode is not added to the neutral flow. It  can also be included, increasing the amplitude of oscillations, but it is not required for the existence of the oscillations. An example of such oscillations shown for a typical Hall thruster parameters (the stationary plasma thruster ``SPT 100''\cite{morozov2000fundamentals}), with the channel length $\SI{3}{cm}$, ionization rate $\beta = \SI{9e-14}{m^3 s^{-1}}$, $v_a = \SI{150}{m/s}$. Atom density at the anode $n_{a0}$ is evaluated from $n_{a0} v_{a}=\dot{m}/Am_{a}$, assuming $\dot{m} = \SI{5}{mgs^{-1}}$, Xenon atom mass, and $A = \SI{12.75 \pi}{cm^2}$. For the simplicity, we assumed the linear ion flow velocity profile $v_i = \left(2.5x - 1.5\right) ~\si{km/s}$ ($x$ is in [\si{cm}]), with the ion back-flow extent $L_b = \SI{0.6}{cm}$. The resulting time and space evolution of ion and atom densities is shown in Fig.~\ref{dens_osc}. The observed frequency is $\SI{12.8}{kHz}$, in the range of the values observed in experiments and simulations for breathing mode.

\begin{figure}[!htbp]

\subfloat[]{\includegraphics[width=0.5\textwidth]{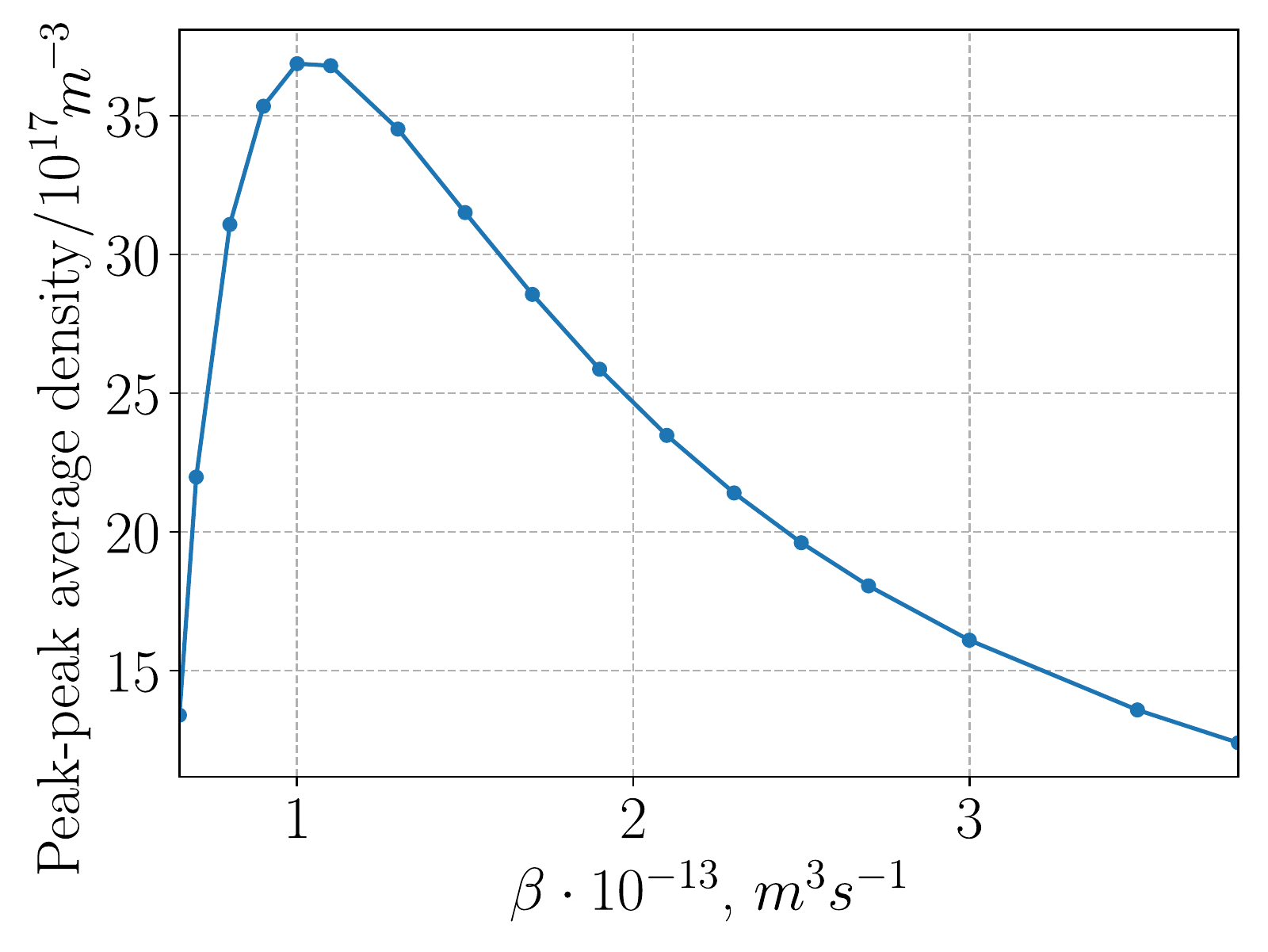}\label{nimax_beta}}
\subfloat[]{\includegraphics[width=0.5\textwidth]{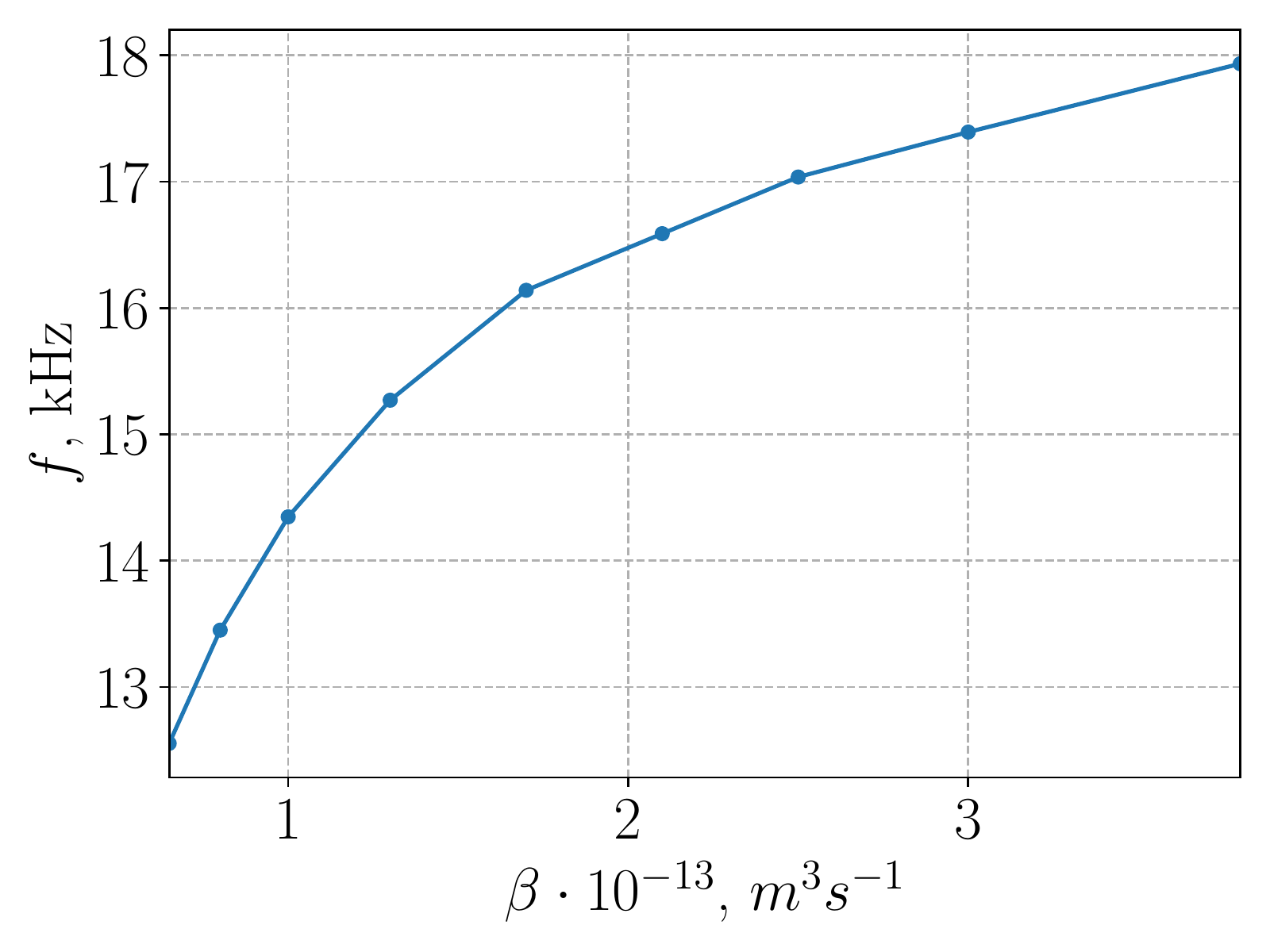}\label{freq_beta}}

\caption{Peak-to-peak values of averaged in space ion density $\langle n_i \rangle_x(t)$ as a function of ionization rate (a). The main frequency component of observed oscillations  as a function of the ionization rate (b).}
\end{figure}

Now, for the same setup with ion velocity profile in the form $v_i = \left(2.5x - 1.5\right) ~\si{km/s}$ ($x$ is in [\si{cm}]), we will show effects of the amplitude of the ionization rate $\beta$, which  is still assumed uniform here for simplicity.   The ionization rate coefficient $\beta$ needs to be sufficiently large to support plasma discharge. Interestingly, in addition to  a threshold $\beta$  value for the discharge to exist, there is also a threshold between stable and oscillatory plasma behaviour. In the oscillatory regime, there is a value of $\beta$ when the amplitude of the oscillations is the largest (note that here we assume that $\beta$ is constant along the channel). It shouldn't be of surprise, as when the values of $\beta$ are too high, the ionization takes place immediately near the left end and so both density profiles have maximum values at the left end. The dependency of the oscillation amplitude on $\beta$ is shown in Fig.~\ref{nimax_beta}, where the peak-to-peak amplitude of the averaged in space density $\langle n_i \rangle_x(t)$ is shown. 


The frequency dependence on $\beta$ is presented in Fig.~\ref{freq_beta}. If the ion back-flow region is very short, i.e.\ $v_i(0) \approx 0$, the ionization mostly will occur at the anode for the same reasons, and there would not be enough spatial separation between high and low atom density points. Dependencies, similar to those shown in Figs.~\ref{nimax_beta}, \ref{freq_beta}, are also observed for different  values of the ion back-low extent $L_b$. They follow the same scaling as presented below in Figs. ~\ref{freq_lb}, \ref{nimax_lb_effect}.





The effect of the ion back-flow region is investigated further for our reduced model by taking the ion velocity profiles with a variable ion back-flow extent. The  Fig.~\ref{backflow_profs} shows the velocity profiles taken as  parabolas specified with the three  points, $v_i(0) = \SI{-1.5}{km/s}$, $v_i(L) = \SI{20}{km/s}$, and $v_i(L_b) = \SI{0}{km/s}$, where $L_b$ is the length of the ion back-flow region, the distance from the anode to the transition point where the ion velocity reverse its sign. The value of $L_b$ was varied, with the system length set to \SI{3}{cm}, the ionization rate $\beta = \SI{5e-14}{m^3/s}$, atom flow velocity $v_a = \SI{150}{m/s}$. Simulations using these velocity profiles reveal that that oscillation frequency scales with the atom fly-by frequency $v_a/L_b$ in the back-flow region, Fig.~\ref{freq_lb}. The oscillations amplitude decreases with decreasing  $L_b$ and the oscillations disappear for  $L_b \to 0$, as shown in Fig.~\ref{nimax_lb_effect}. Also, no oscillations were observed for monotonically increasing ion velocity profiles with $v_i(0) \ge 0$.

\begin{figure}[!htbp]
\centering
\subfloat[]{\includegraphics[width=0.5\textwidth]{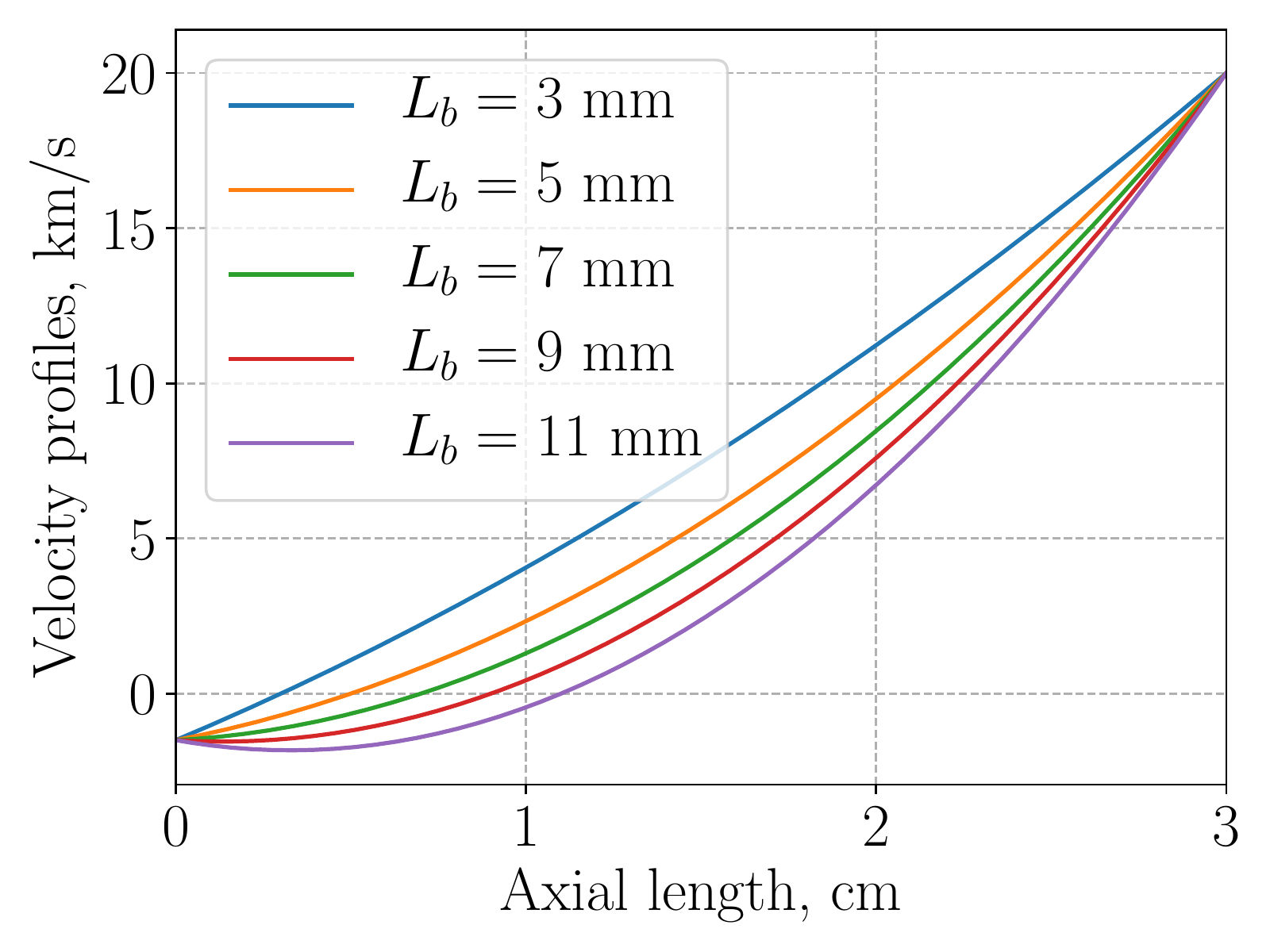}\label{backflow_profs}}
\subfloat[]{\includegraphics[width=0.5\textwidth]{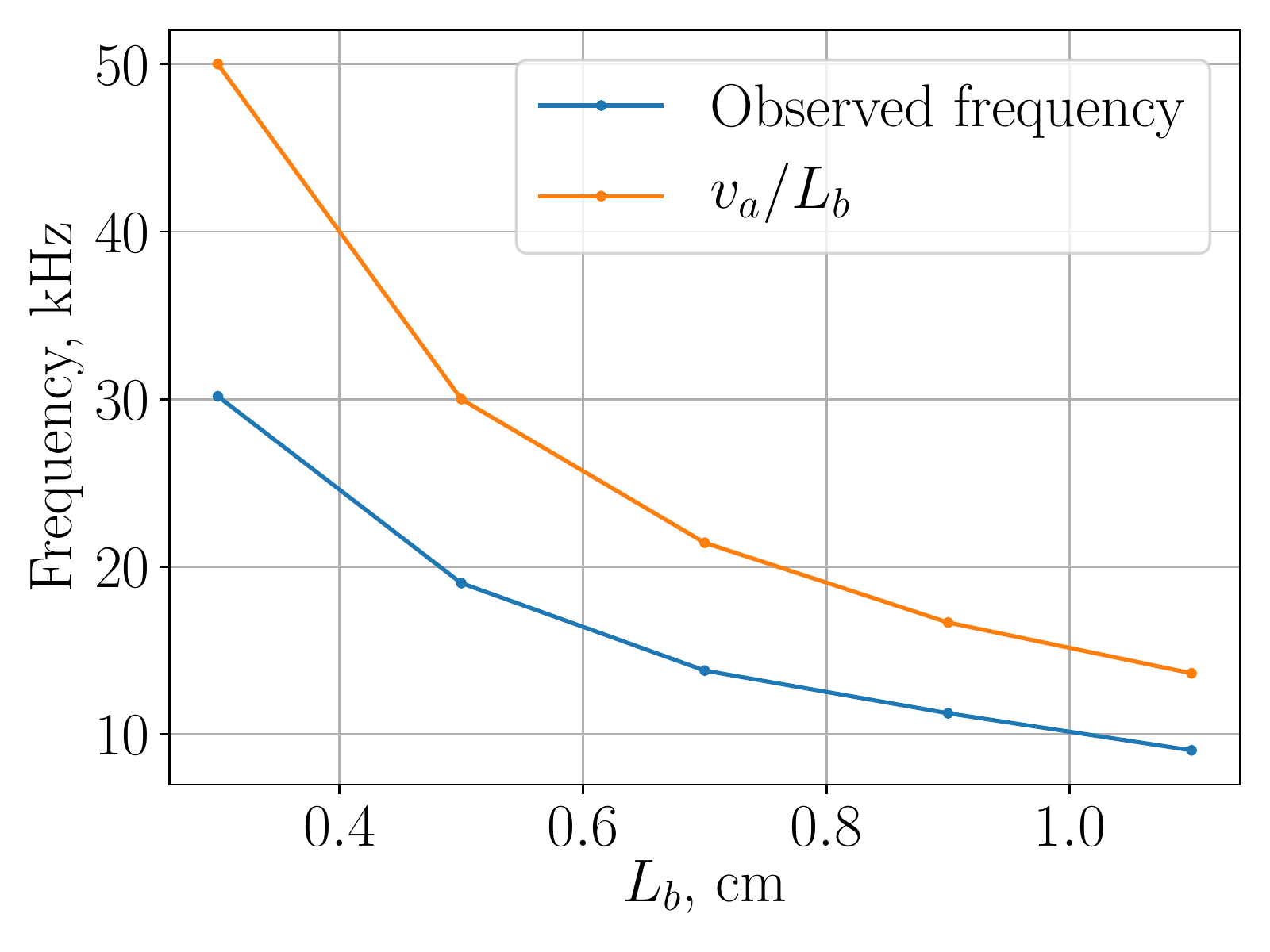}\label{freq_lb}}
\caption{Velocity profiles (a) for a variable ion back-flow region $L_b$ while the exhaust velocity is fixed to $\SI{20}{km/s}$ (\SI{300}{eV} is a typical ion energy in the exhaust plume of ``SPT100''\cite{morozov2000fundamentals}). The corresponding frequency (b) as a function of $L_b$ in compare with neutral flyby frequency of $v_a/L_b$ in the backflow region.}
\end{figure}

\begin{figure}[!htbp]
\centering
\subfloat{\includegraphics[width=0.5\textwidth]{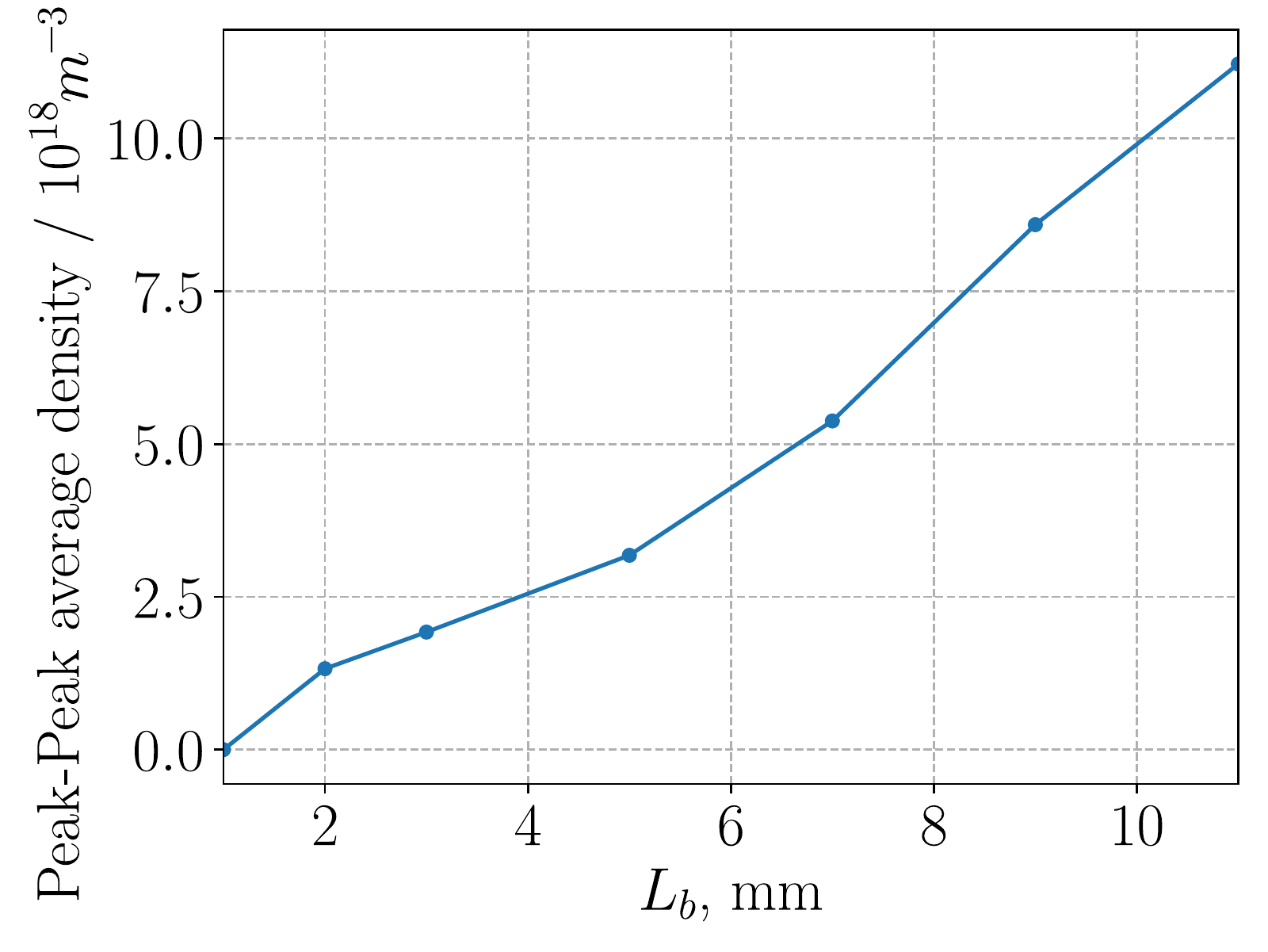}}
\caption{Peak-to-peak values of averaged in space ion density $\langle n_i\rangle_x(t)$ as a function of the ion back-flow region $L_b$. Oscillations disappear as $L_b=\SI{1}{mm}$ goes to zero.}
\label{nimax_lb_effect}
\end{figure}

\begin{figure}[!htbp]
\centering
\subfloat[]{\includegraphics[width=0.5\textwidth]{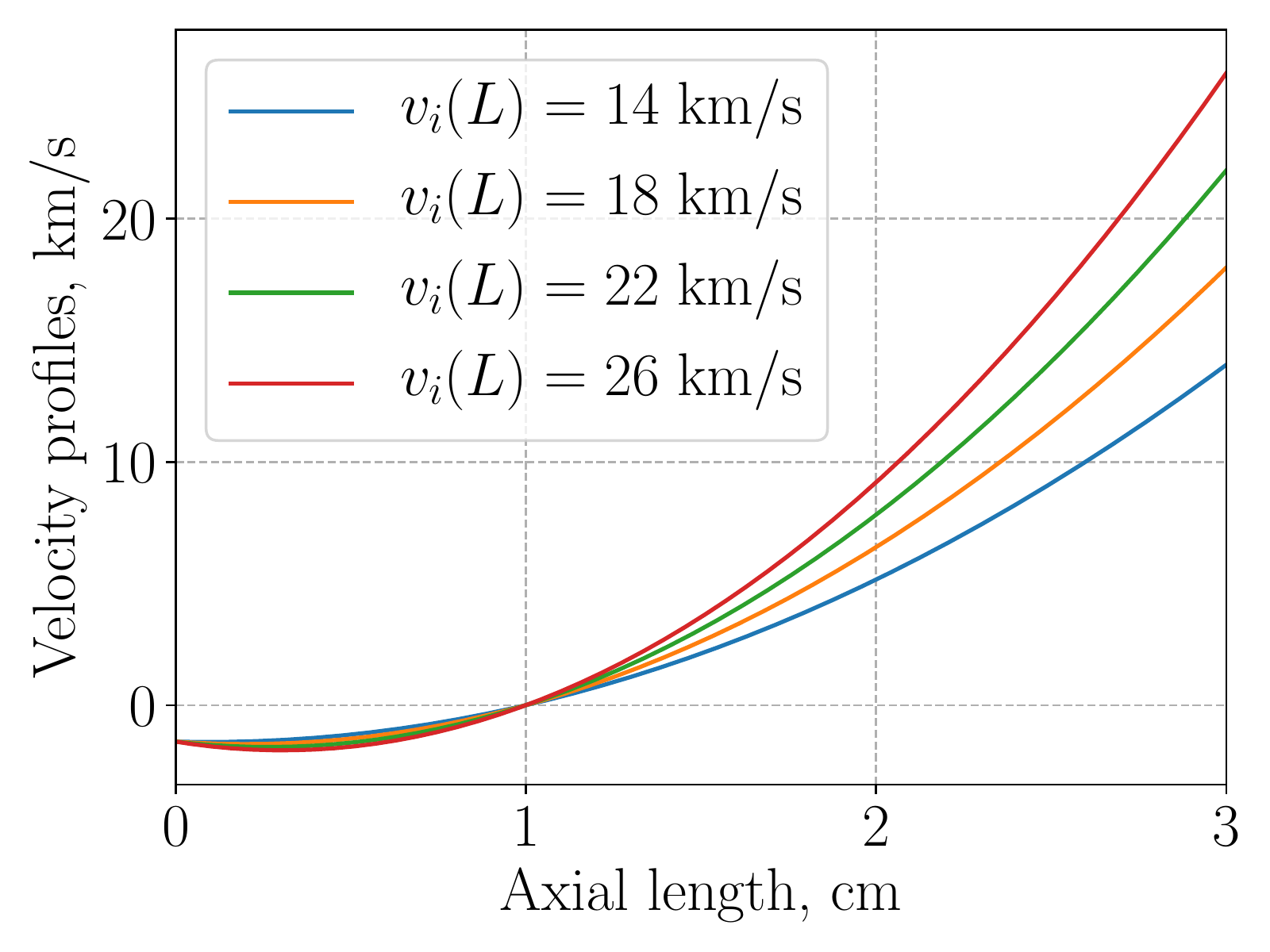}\label{vil_profs}}
\subfloat[]{\includegraphics[width=0.5\textwidth]{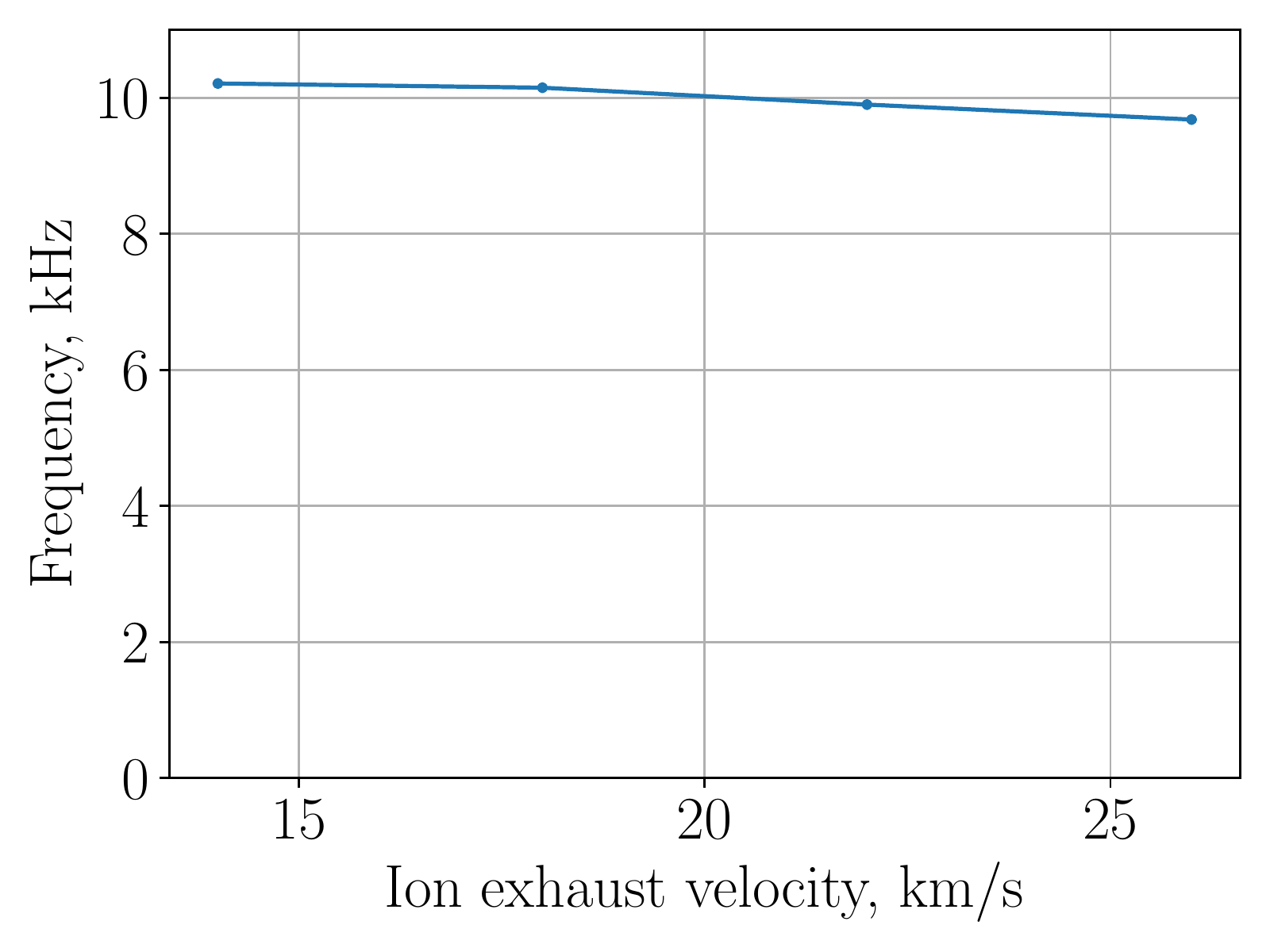}\label{freq_vil}}
\caption{Velocity profiles (a) and   corresponding frequency (b) as a function of $v_i(L)$, for the variable ion exhaust velocity $v_i(L)$ while keeping the back-flow region $L_b$ constant. }
\end{figure}

Zero-dimensional predator-prey models predict that the frequency depends on the exhaust ion velocity. Here, we show that for our model this effect is very small and, mostly, the back-flow region ``defines'' the frequency. Fig.~\ref{vil_profs} shows the family of the ion velocity profiles, constructed in a similar manner (by a parabola) but keeping the back-flow length the same (\SI{1}{cm}) and  varying $v_i(L) = (14,18,22,26)~\si{km/s}$. The resulting frequency for all these profiles was found to be close to 10 kHz, Fig.~\ref{freq_vil}, predicted by the previous example in Fig.~\ref{freq_lb} with the  $L_b = \SI{1}{cm}$.

The mechanism of the oscillations can be considered  in  three main stages, starting from a stationary state  in Fig.~\ref{stnr_sols1}, but with the ion velocity back-flow region. In the first stage we observe enhanced ionization process in the region of plasma stagnation ($v_i \approx 0$). The ionization process in this stage may be approximated as $\dot{n}_i = \beta n_i n_a$, $\dot{n}_a = -\beta n_i n_a$ (where $n_i, n_a$ are some average plasma and atom densities, respectively, in that region). In this stage,  effects of  atom inflow and ion outflow are relatively small here. The ionization proceeds with the exponential rise plasma density $n_i \sim \exp{\left(\beta n_a t\right)}$ and depletion of atom density. In the second stage,  the newly created plasma peak splits and partially moves to the anode (via back-flow) wiping  down the rest of atoms near the anode. Finally, in the third stage, the system enters the period of ``regeneration'' with both plasma and atom densities low, while atoms slowly fill the region near the anode. The atom density remains higher than the ion density, and  the ion density  balance is predominantly follows the equation $\dot{n}_i = \beta n_i n_a - \gamma_{loss} n_i$, where $\gamma_{loss} n_i$ represents the ion advection losses via the back-flow. In this quasistationary state, there is no exponential rise of ionization but slow inflow  of atoms (to the right) filling this region toward the stagnation region. The third stage is also the slowest  and  defines the breathing mode period, hence the backflow lengths defines the period (Fig.~\ref{freq_lb}). Additionally, if the recombination at the anode is included, it increases a number of atoms injected to the system for the stage three, and it increases the amplitude of the oscillations.


\section{Reduced vs full self-consistent model}\label{sec_comparison}

In this section we compare the reduced and full models for the breathing  modes, and demonstrate that our  simple reduced model, Eqs.~(\ref{na_cont},\ref{ni_cont}), reproduces well the results from  the full self-consistent time-dependent model \cite{smolyakov2019theory}. First, we give a brief description of the full fluid self-consistent model for low-frequency ionization oscillations. 

\subsection{Full self-consistent model}
\begin{figure}[htbp]
\centering
\subfloat[]{\includegraphics[width=0.5\textwidth]{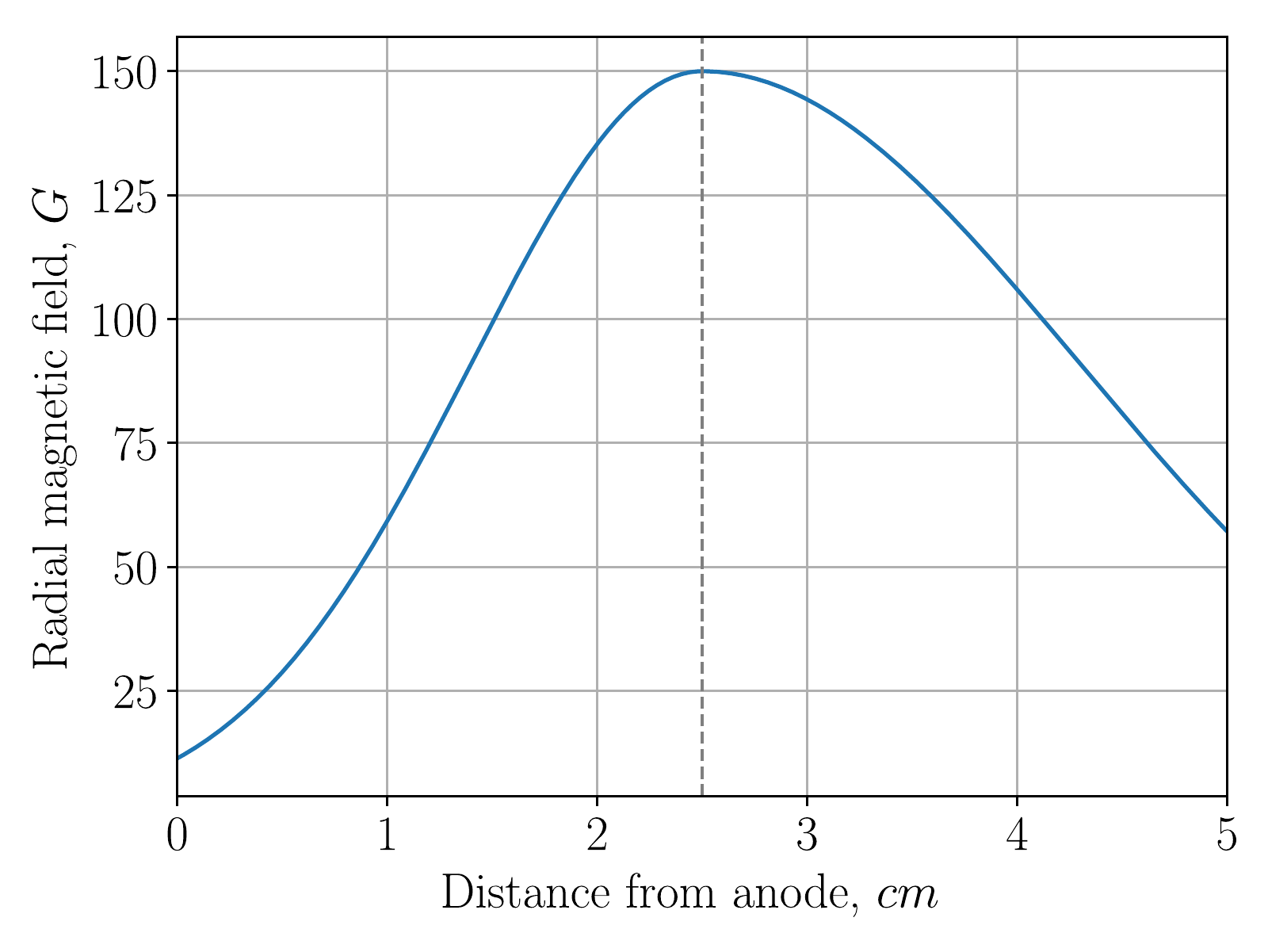}\label{b_profile}}
\subfloat[]{\includegraphics[width=0.5\textwidth]{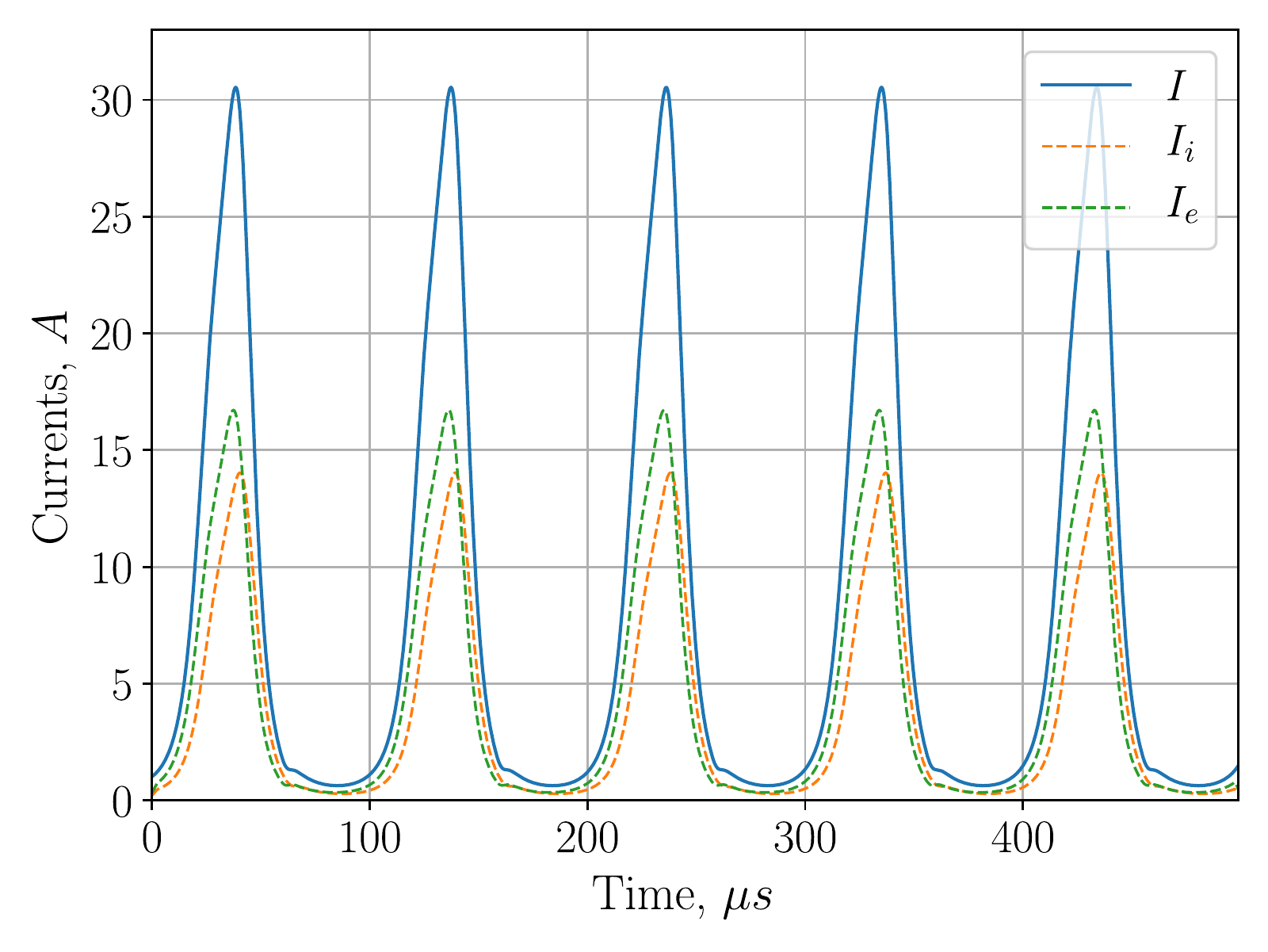}\label{typ_full_current}}
\caption{The magnetic field profile (a) used in full fluid and hybrid simulations, with the channel exit located \SI{2.5}{cm} from anode (dashed line). Low-frequency oscillations in time (b) of total current $I$, ion current $I_i$ (at $x=\SI{5}{cm}$), and electron current $I_e$ (at $x=\SI{5}{cm}$) in the full fluid model using specified parameters.}
\end{figure}

The full model of low-frequency axial plasma dynamics in Hall thruster is considered in electrostatic and quasineutral approximation for three species: neutrals, ions, and electrons.  Our model is fully fluid both for the ion, neutral, and electron components. The electron equations in general are similar to those used in Refs. \cite{boeuf1998low, barral2009low}. The full fluid model consists of time-dependent PDE equations for neutral atom density $n_a$, plasma density $n$ (ion and electron), ion flow velocity $v_i$, and electron temperature $T_e$:
\begin{eqnarray}
&& \frac{\partial n_a}{\partial t} + v_a \frac{\partial n_a}{\partial x} = -\beta n_a n, \label{sys:na} \\
&& \frac{\partial n}{\partial t} + \frac{\partial}{\partial x}\left(n v_i\right) = \beta n_a n, \label{sys:ni} \\
&& \frac{\partial v_i}{\partial t} + v_i\frac{\partial v_i}{\partial x} = \frac{e}{m_i}E + \beta n_a \left(v_a - v_i\right), \label{sys:vi} \\
&& \frac32 \frac{\partial }{\partial t} \left(nT_e\right) + \frac52 \frac{\partial }{\partial x}\left(nv_{ex} T_e \right) + \frac{\partial q_e}{\partial x} = nv_{ex}E - nn_a \mathrm{K} - n \mathrm{W}, \label{sys:te}
\end{eqnarray}
where the electric field $E$ is obtained from the electron momentum balance equation
\begin{equation}\label{ve}
v_{ex} = -\mu_e E - \frac{\mu_e}{n_e} \frac{\partial (nT_e)}{\partial x},
\end{equation}
obtained with neglected electron inertia, where $\mu_e$ is the electron mobility across the magnetic field (described below). 
Quasineutrality condition, along with the constraint $\int_L E dx = U_0$ ($U_0$ is the applied potential) leads to the total current $J_T = en \left( v_i - v_e\right)$ 
\begin{equation} \label{jt-nodiff}
J_T = \dfrac{U_0 + \bigintss_0^L \left( \dfrac{v_i}{\mu_e}  + \dfrac{1}{n} \dfrac{\partial p_e}{\partial x} \right) dx} {\bigintss_0^L \dfrac{dx}{en\mu_e}},
\end{equation} 
which is spatially uniform but  may oscillate in time.
Other quantities in the system~(\ref{sys:na}-\ref{sys:te}) are the constant atom flow velocity $v_a$, the ionization rate coefficient $\beta$ (obtained with BOLSIG+\cite{hagelaar2005solving} for Maxwellian EEDF using SIGLO database\cite{siglo}), the elementary charge $e$, the electron mass $m_e$, the ion (Xenon) mass $m_i = \SI{131.293}{amu}$, the anomalous energy loss coefficient\cite{boeuf1998low} $\mathrm{W}$, the collisional energy loss coefficient $\mathrm{K}$ (also via BOLSIG+), and the electron heat flux given by 
\begin{equation}
q_e = -\frac{5}{2} \mu_e n T_e \frac{\partial T_e}{\partial x}.
\end{equation} 
The electron transport across magnetic field is described in the form of the magnetized mobility
\begin{equation}\label{mu_cl}
\mu_e = \frac{e}{m_e \nu_m} \frac{1}{1+\omega_{ce}^2/\nu_m^2},
\end{equation}
where $\nu_m$ the total electron momentum exchange collision frequency, $\omega_{ce} = eB/m_e$ is the electron cyclotron frequency. In this model $\nu_m$ is represented in the form as was adopted in Ref. \cite{hagelaar2004modelling},
\begin{equation}\label{nue}
\nu_{m} = \nu_{en} + \nu_{walls} + \nu_{B},
\end{equation}
where the electron-neutral collision frequency $\nu_{en}$, electron-wall collision frequency $\nu_{walls}$, and anomalous Bohm frequency $\nu_{B}$ are given with:
\begin{eqnarray}
\nu_{en} = k_m n_a, \\
\nu_{walls} = \alpha \SI{e7}{[s^{-1}]}, \label{nuw}\\
\nu_{B} = \left(\beta_a/16\right) eB/m_e. \label{nub}
\end{eqnarray}
where $k_m = \SI{2.5e13}{m^{-3} s^{-1}}$, $\alpha$ and $\beta_a$ are adjusting constants. The profile of external magnetic field $B$ is shown if Fig.~\ref{b_profile}, with the channel's exit in the peak of magnetic field intensity, given by $B = B_0 \exp{\left(-(x-l)^2/2 \delta_B^2 \right)}$, where $l=\SI{2.5}{cm}$ is the channel length and $\delta_B$ defines a magnetic field profile width. This model use different parameters inside ($x < l$) and outside ($x \ge l$) the channel \cite{hagelaar2003role,hagelaar2004modelling,hagelaar2007modelling}, the near wall conductivity contribution~(\ref{nuw}) $\alpha_{in} = 0.2, \ \alpha_{out} = 0$, and the anomalous contribution~(\ref{nub}) is set to $\beta_{a,in} = 0.1$, $\beta_{a,out} = 1$. The anomalous electron energy loss coefficient\cite{boeuf1998low} $\mathrm{W}$ is modeled as
\begin{equation}\label{el_an_loss}
\mathrm{W} = \nu_{\varepsilon} \varepsilon \exp{\left(-U/\varepsilon\right)},
\end{equation}
where $\varepsilon = 3/2 T_e$, $U = \SI{20}{V}$, and $\nu_{\varepsilon}$ is electron energy anomalous loss coefficient. A constant mass flow rate $\dot{m}$ determines the value of $n_a$ at the boundary together with the recombination of plasma that flows to the anode, hence the boundary condition:
\begin{equation}\label{nn_backflow}
n_a(0) = \frac{\dot{m}}{m_i A v_a} - \frac{nv_i(0)}{v_a},
\end{equation}
where $A$ is the anode surface area of a thruster. 
Bohm type condition for ion velocity can be imposed at the anode $v_i(0) = -b_v \sqrt{T_e/m_i}$, where $b_v=0\text{--}1$ is the Bohm velocity factor which can be varied. Both anode and cathode electron temperature are fixed with $T_e(0) = T_e(L) =  2~\si{eV}$. All other boundary conditions are free (spatial second derivative is zero).

Following the LANDMARK benchmark\cite{landmark} Test Case 3, we use the parameters that result in bulk low-frequency oscillations, with the electron energy anomalous loss coefficient inside the channel $\nu_{\varepsilon,\text{in}} = 10^7~\si{s^{-1}}$ and outside $\nu_{\varepsilon,\text{in}} = 0.4 \cdot 10^7~\si{s^{-1}}$.
The resulting plasma currents evolution shown in Fig.~\ref{typ_full_current}, assuming the typical inner and outer radius of a Hall thruster (``SPT 100''\cite{morozov2000fundamentals}) $R_1 = \SI{3.5}{cm}, R_2 = \SI{5}{cm}$, respectively. The input atom mass flow used in the simulation is $\dot{m} = \SI{5}{mgs^{-1}}$ corresponds to $I = \dot{m}q_e /m_i = \SI{3.67}{A}$. The average ion current (to the exit plane) is \SI{3.68}{A} which is consistent with the mass flow rate. The electron current (to the exit plane) is of the same order (or slightly larger). Other simulation parameters for this test case are: $\delta_{B,in} = \SI{1.1}{cm}$, $\delta_{B,out} = \SI{1.8}{cm}$, $A = \pi (R_2^2 - R_1^2)$, $b_v = 1$.

The self-consistent fluid model used here is essentially based on the formulations and parameters suggested in Refs. \cite{boeuf1998low,hagelaar2002two,hagelaar2004modelling,hagelaar2007modelling}, also see the LANDMARK benchmark\cite{landmark}. The fluid model has been compared  against the hybrid model \cite{hagelaar2004modelling, hallis}; both models demonstrate similar results \cite{smolyakov2019theory}.

\subsection{Comparison of the reduced model with  predictions of the  full self-consistent model}
\begin{figure}[htbp]
\centering
\subfloat[]{\includegraphics[width=0.46\textwidth]{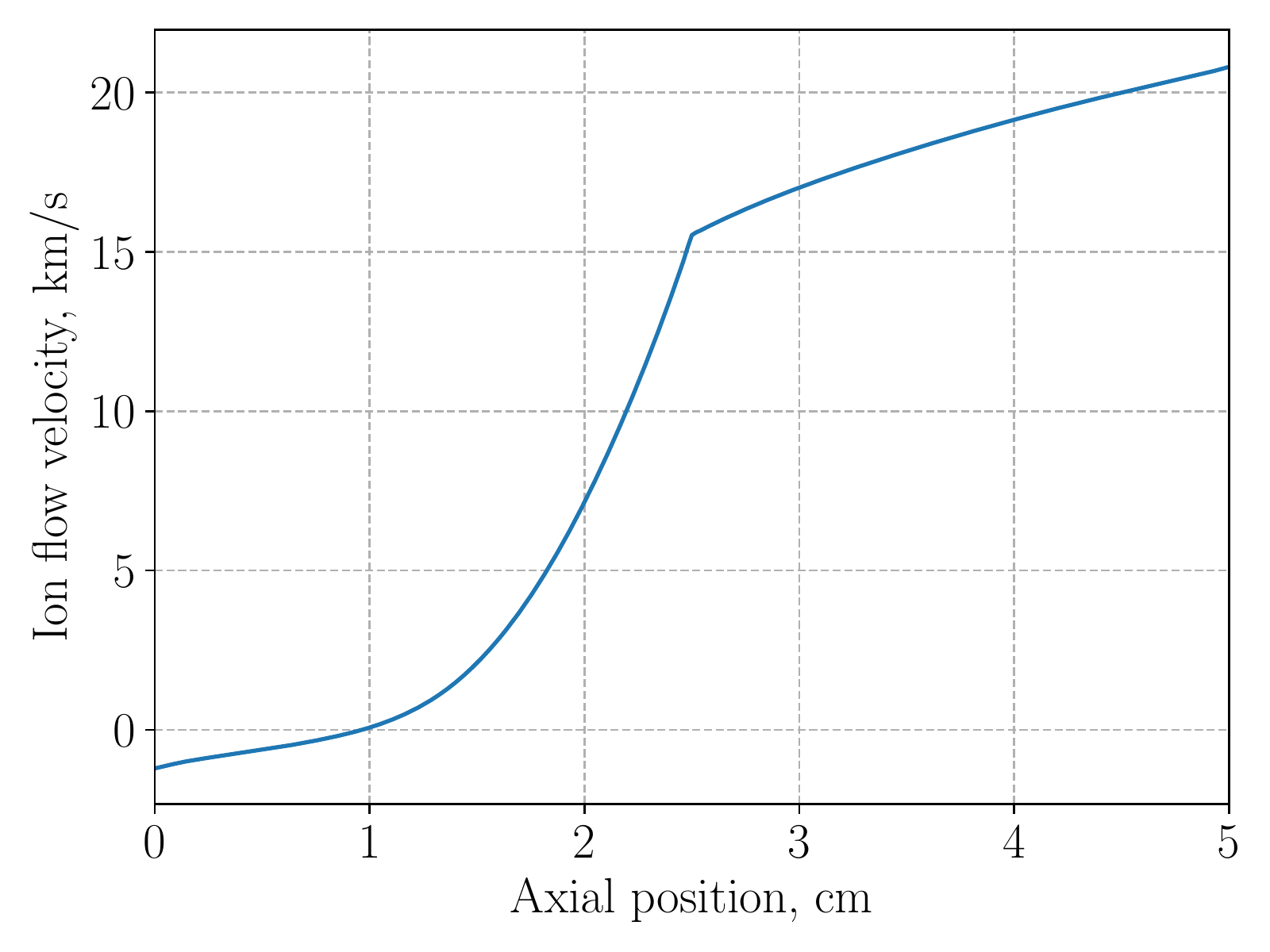}\label{vicase2}}
\subfloat[]{\includegraphics[width=0.46\textwidth]{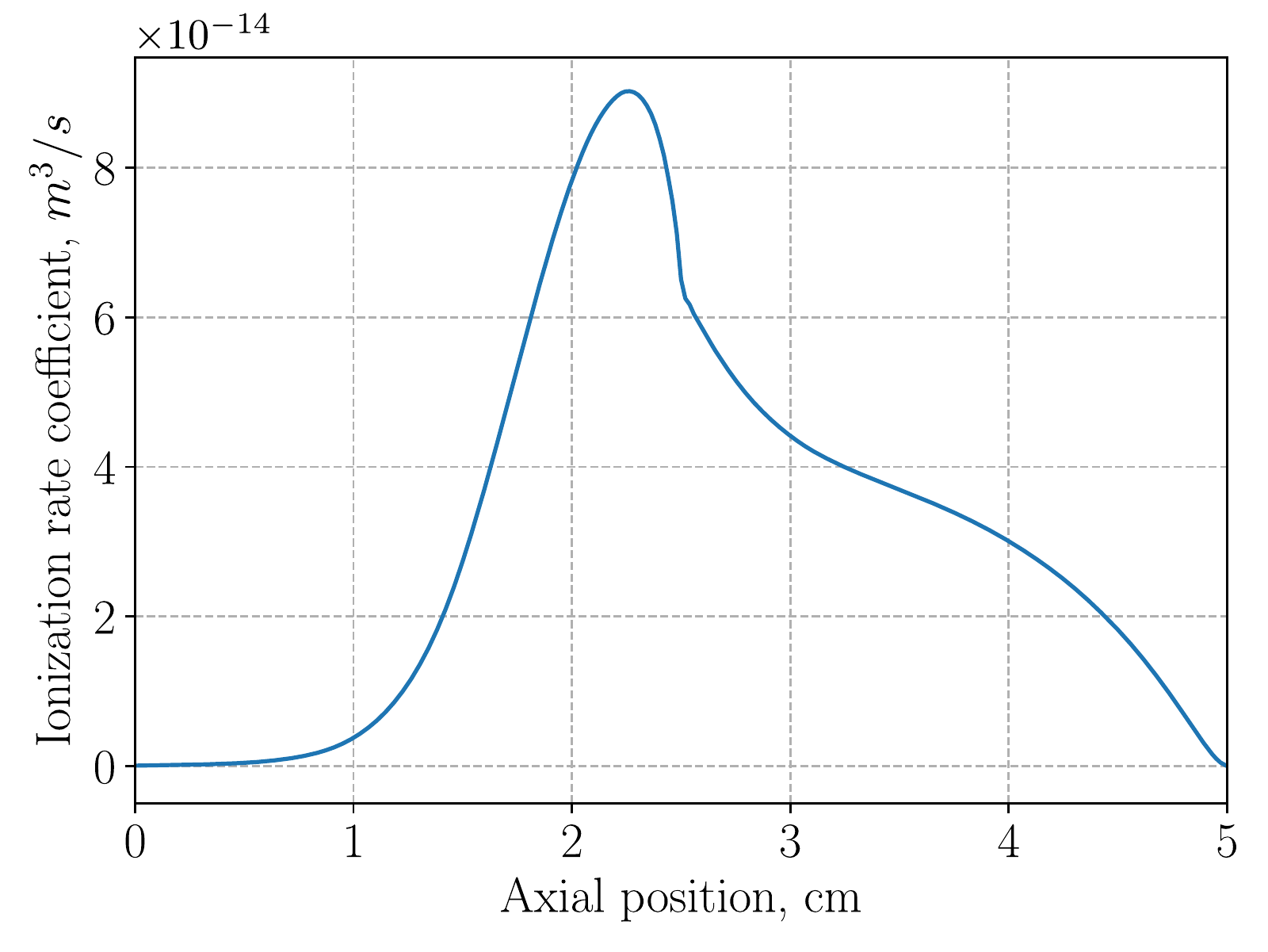}\label{betacase2}}
\caption{Ion velocity (a) and ionization rate (b) coefficient profiles obtained from the full fluid model as time-averaged over a few oscillation periods. They are used in the reduced model as fixed profiles in time.}
\end{figure}

We compare now results of the full fluid model (in the regime described above), with the reduced model. In the full model the ion velocity profile $v_i$ and the ionization rate $\beta$ profile are self-consistent variables, and for the reduced model we take averaged in time profiles of these variables, Figs.~\ref{vicase2}, \ref{betacase2}. We also include the ion back-flow recombination to the atom boundary condition at the anode in the reduced model, described by Eq.~(\ref{nn_backflow}), in agreement with the full model.

\begin{figure}[htbp]
\subfloat[$n_a$ (Full)]{\includegraphics[width=0.33\textwidth]{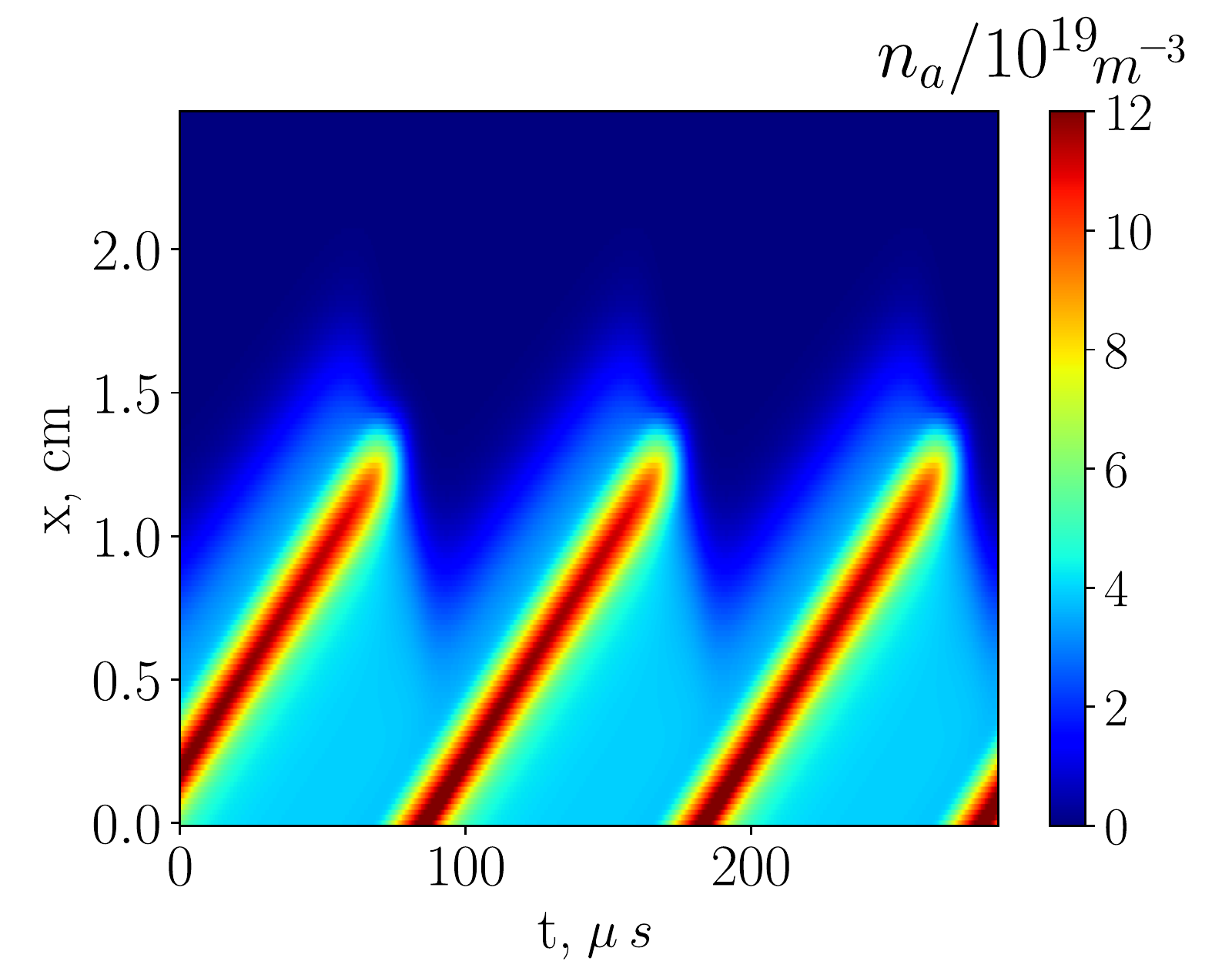}\label{full-na}}
\subfloat[$n_a$ (Reduced)]{\includegraphics[width=0.33\textwidth]{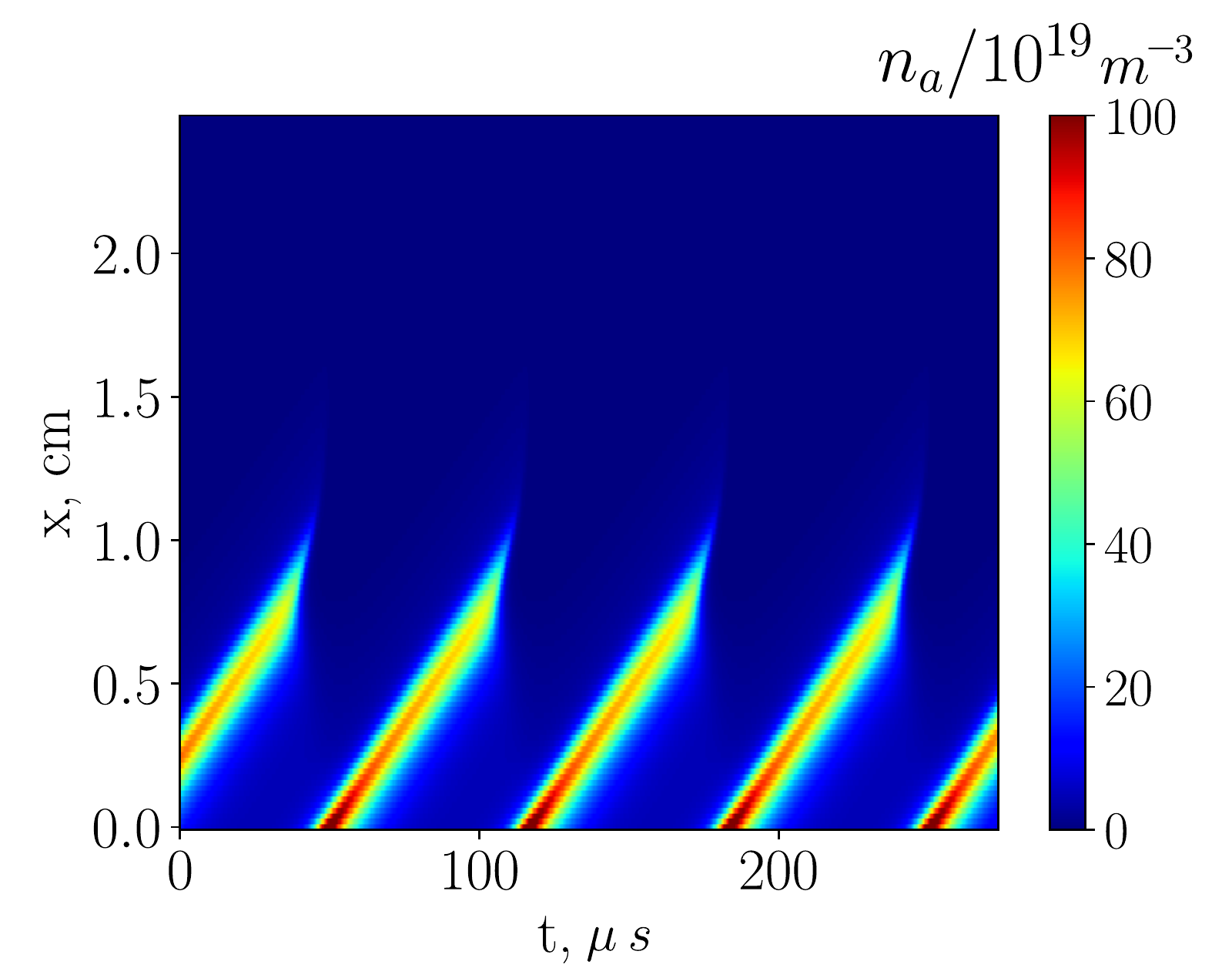}\label{reduced-na}}
\subfloat[$n_a$ (Reduced, lower $\beta$)]{\includegraphics[width=0.33\textwidth]{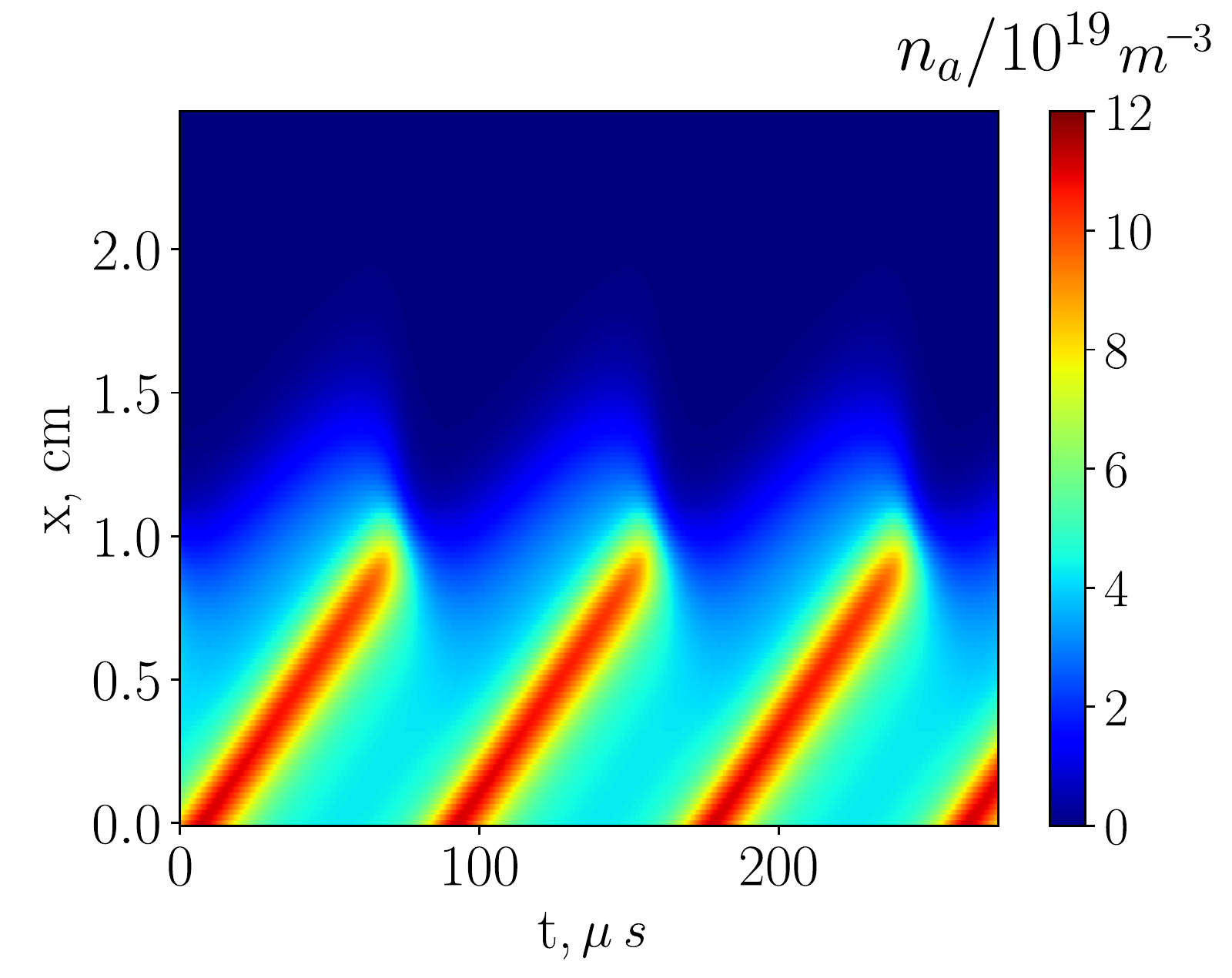}\label{reduced-na2}}

\subfloat[$n_i$ (Full)]{\includegraphics[width=0.33\textwidth]{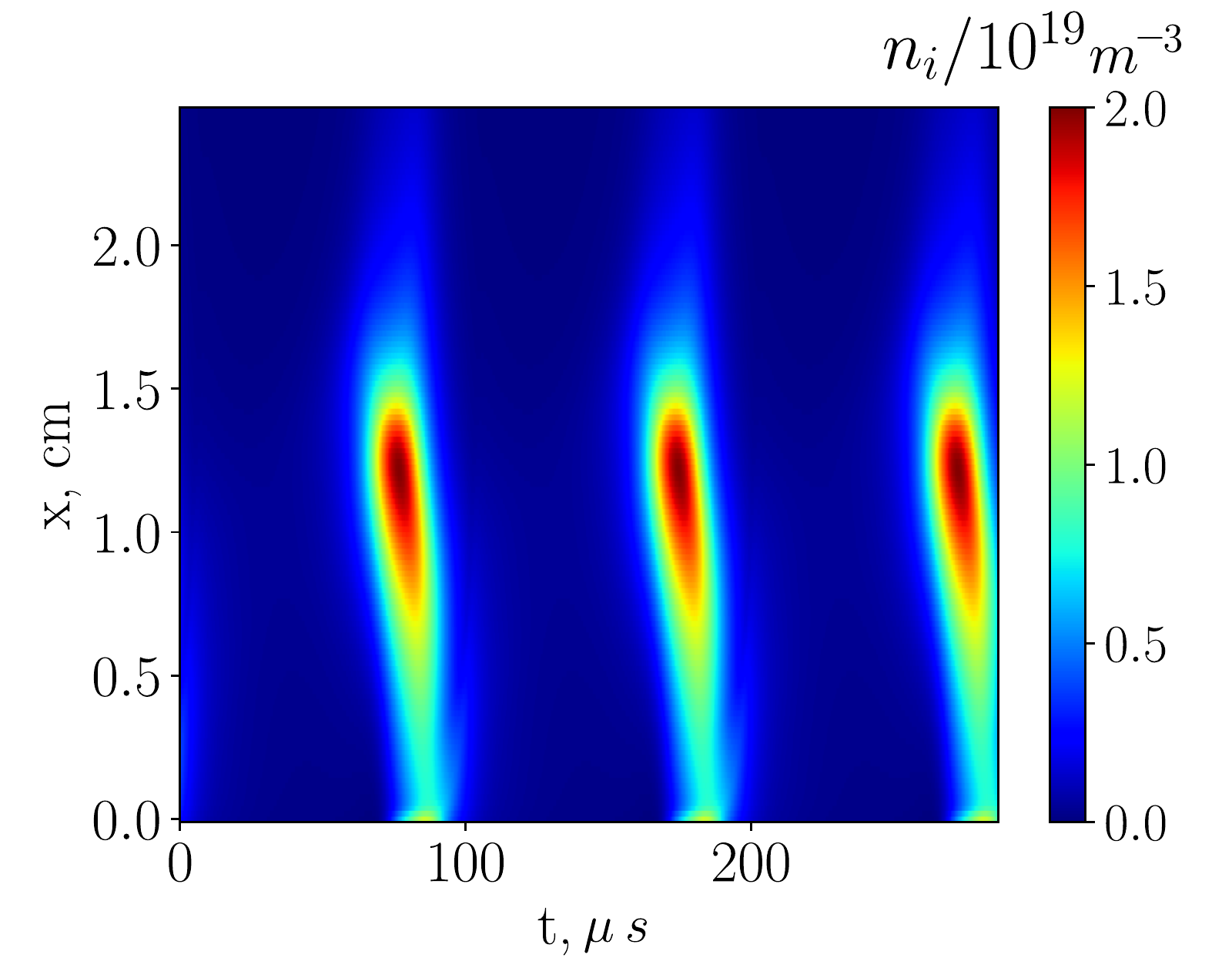}\label{full-ni}}
\subfloat[$n_i$ (Reduced)]{\includegraphics[width=0.33\textwidth]{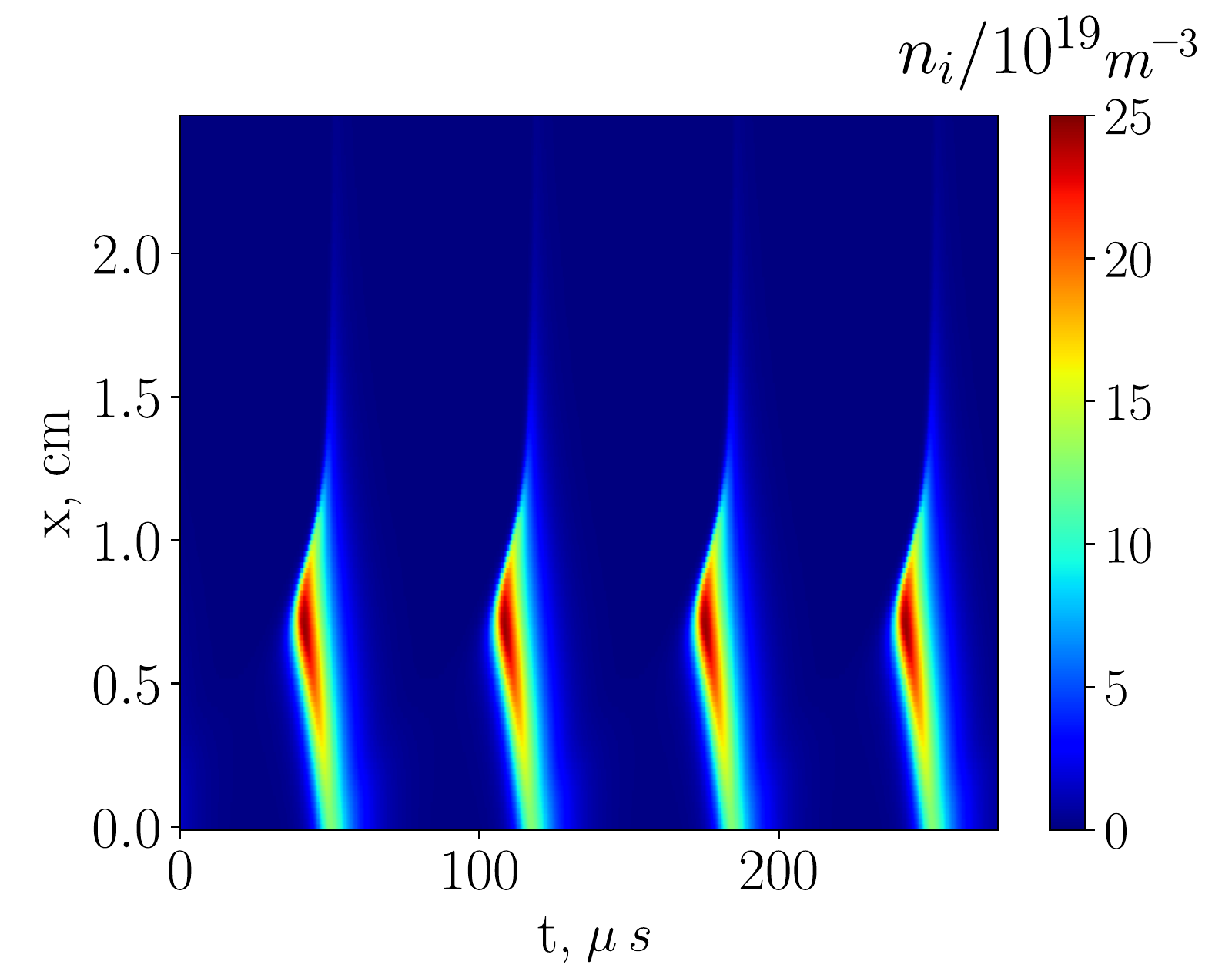}\label{reduced-ni}}
\subfloat[$n_i$ (Reduced, lower $\beta$)]{\includegraphics[width=0.33\textwidth]{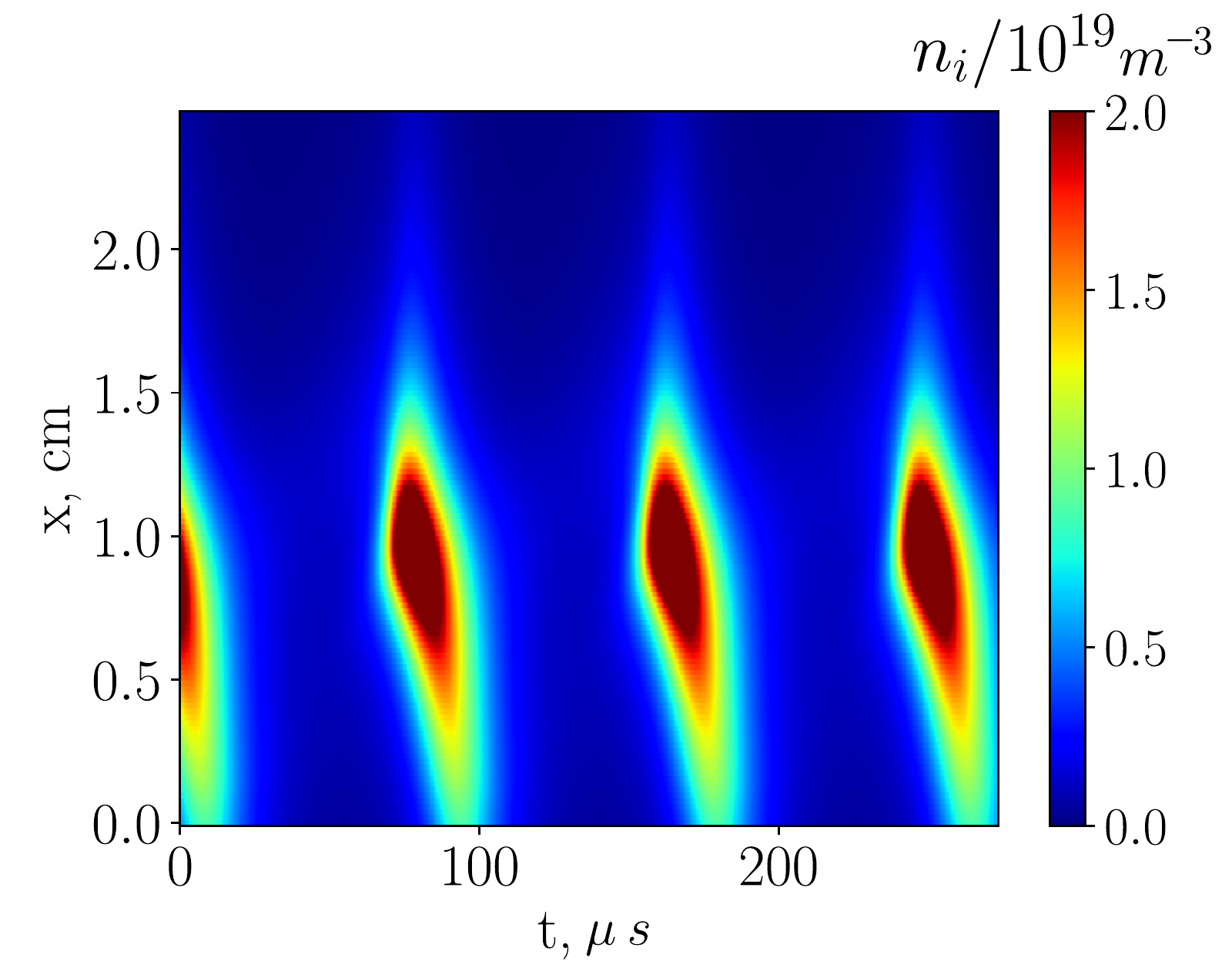}\label{reduced-ni2}}

\subfloat[Source (Full)]{\includegraphics[width=0.33\textwidth]{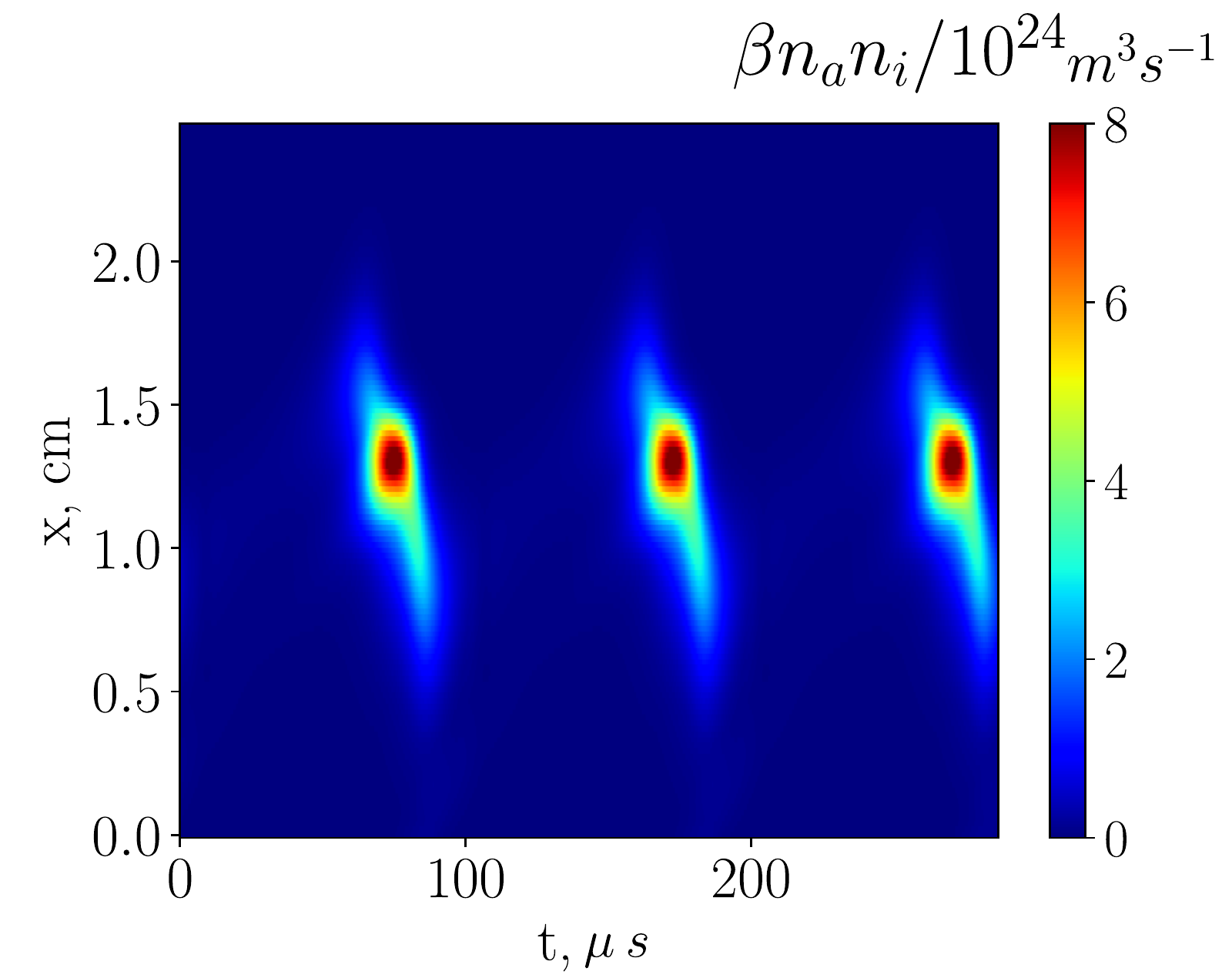}\label{full-s}}
\subfloat[Source (Reduced)]{\includegraphics[width=0.33\textwidth]{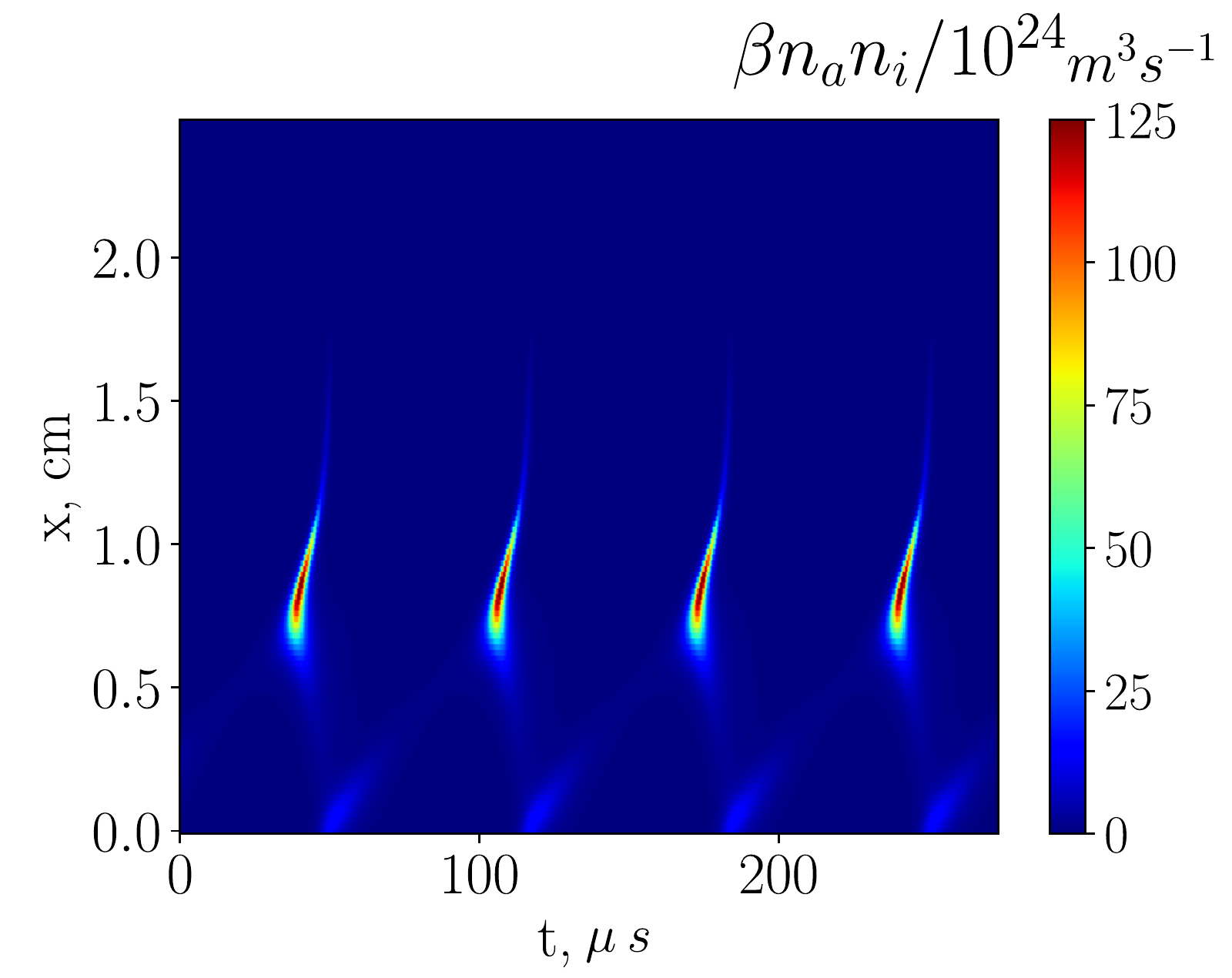}\label{reduced-s}}
\subfloat[Source (Reduced, lower $\beta$)]{\includegraphics[width=0.33\textwidth]{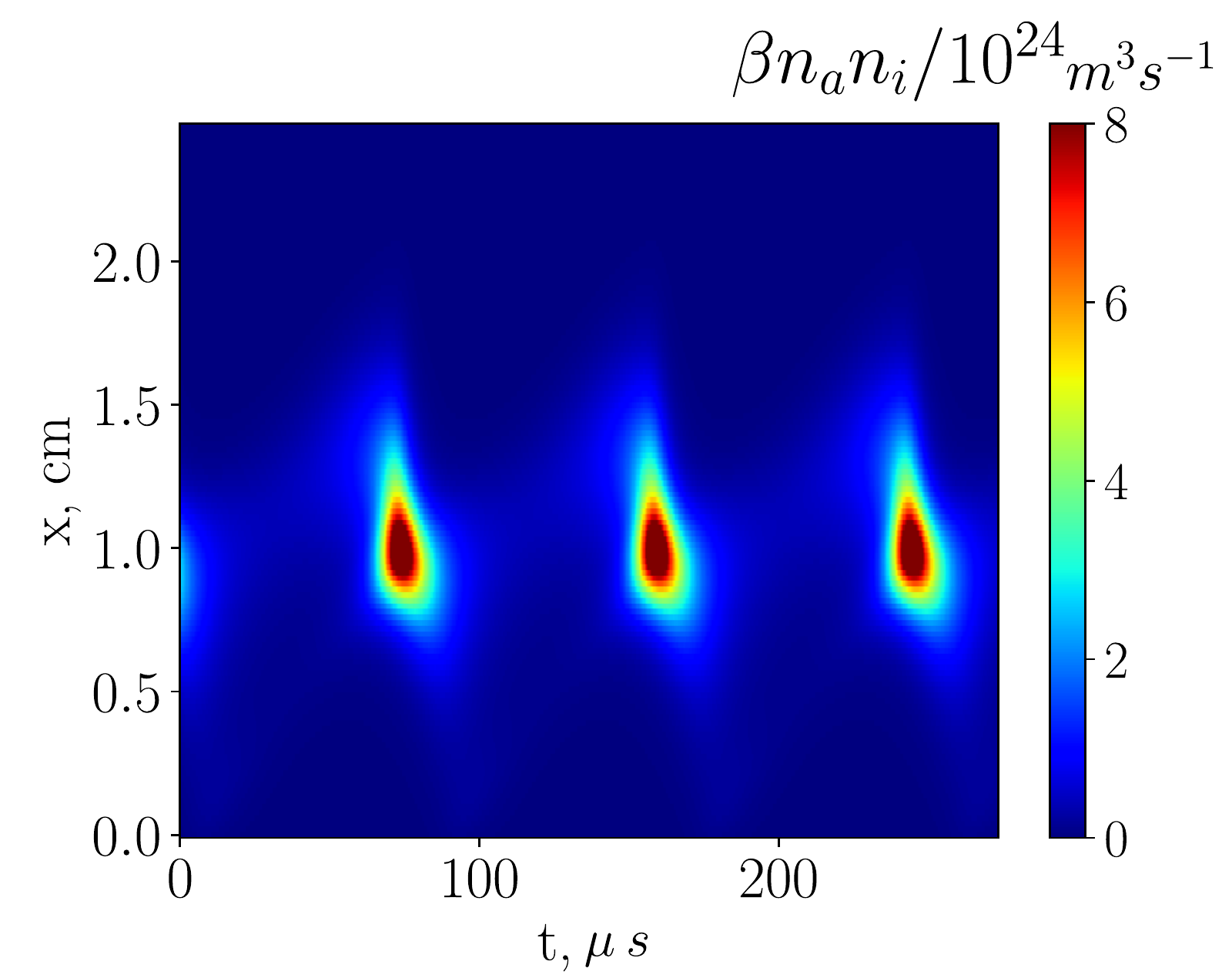}\label{reduced-s2}}
\caption{Neutral density (a,b,c), ion density (d,e,f), and ionization source $\beta n_i n_a$ (g,h,i) for the full fluid model, the reduced model, and the reduced model with lower values of $\beta$ (see captions). Density values are normalized to $\SI{e19}{m^{-3}}$, and source values to $\SI{e24}{m^{-3}s^{-1}}$. The spatial domain is limited to channel region only, $\SI{2.5}{cm}$ from the anode.}
\end{figure}

Time evolution of $n_a$, $n_i$, and the source $S = \beta n_a n_i$, are shown in Figs.~\ref{full-na}, \ref{full-ni}, \ref{full-s}, respectively. The same quantities from the reduced model are shown in Figs.~\ref{reduced-na}, \ref{reduced-ni}, \ref{reduced-s}. It is seen that the reduced model  results in oscillatory behaviour, qualitatively reproducing the breathing mode oscillations, but with much higher amplitudes (about one order) of evolving variables. The observed natural frequency in this model is also higher, 14.9 kHz, compared to 10.2 kHz in the full model.

The observed discrepancy can be explained by the absence of the electron temperature evolution and lack of self-consistent ionization rate dynamics in the reduced model which is known to affect the breathing mode characteristics, e.g.\ Ref.~\onlinecite{Wei2012PoP}. One observation is that the average position of the source term $S$ in our reduced model is closer to the anode (Fig.~\ref{reduced-s}) (compared with the full model) that results in higher frequency. Higher amplitude may also be explained by the shift of the ionization source toward the region with higher ion back-flow velocity so that  more ions return to the anode and recombine enhancing the positive feedback loop.  We found that simply lowering the values of the ionization rate, precisely by taking $0.82 \beta$ of the original averaged $\beta$ profile (shown in Fig.~\ref{betacase2}) agrees with the full model much closer,  see Figs.~\ref{reduced-na2}, \ref{reduced-ni2}, \ref{reduced-s2}, with the oscillation frequency of 11.8 kHz.
While our intention is not really to achieve full quantitative agreement, this exercise shows the sensitivity to the temperature effects. It is also possible that the strong nonlinear dependence of the ionization rate on temperature may suggest some weighting average for the ionization rate.  We leave the investigation of the proper averaging technique for future work.

\begin{figure}[htbp]
\centering
\subfloat{\includegraphics[width=0.7\textwidth]{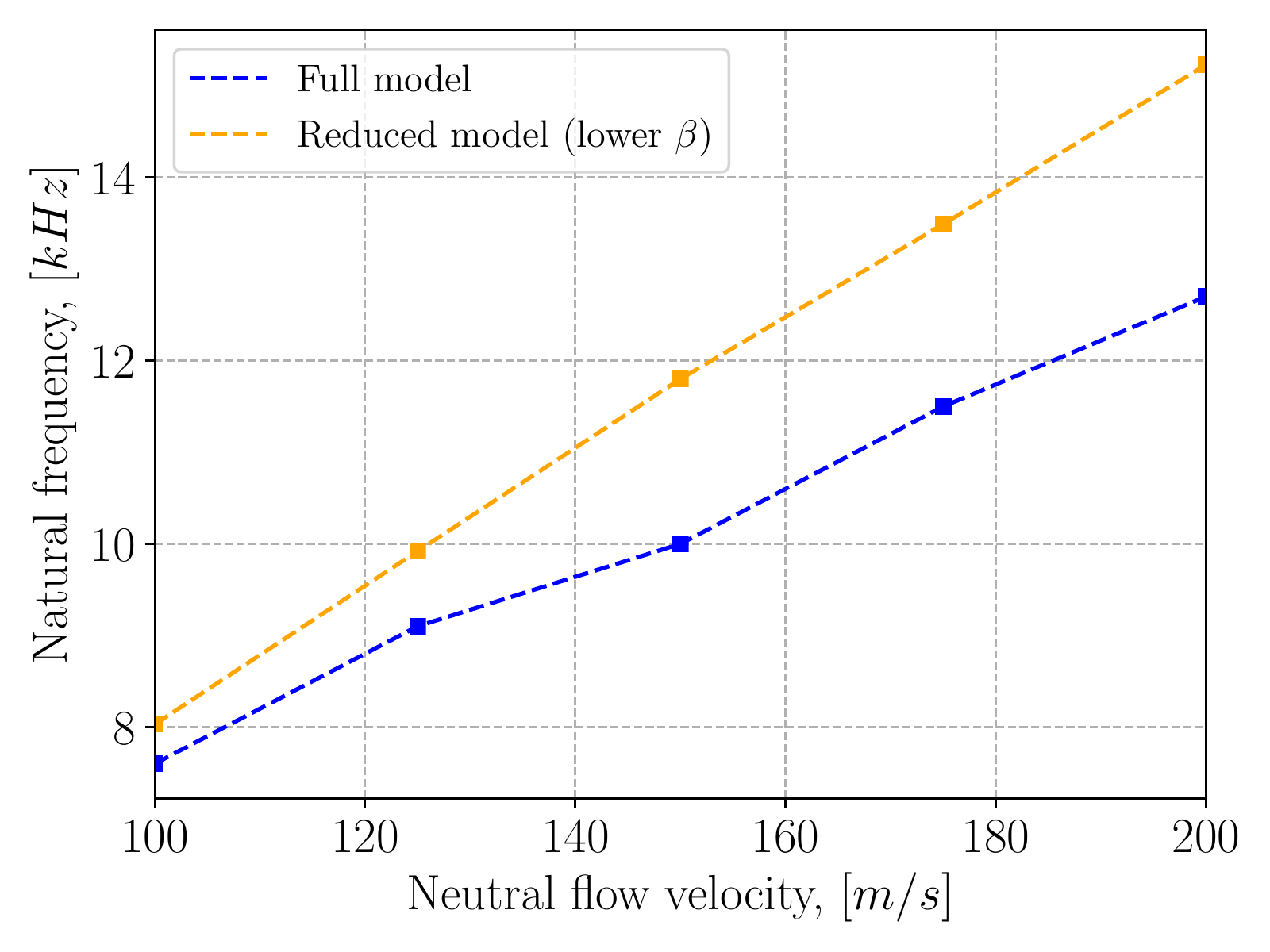}}
\caption{Oscillation frequency as a function of neutral velocity in full and reduced (with lower $\beta$) fluid models.}
\label{va_effect}
\end{figure}

\begin{figure}[htbp]
\centering
\subfloat[]{\includegraphics[width=0.5\textwidth]{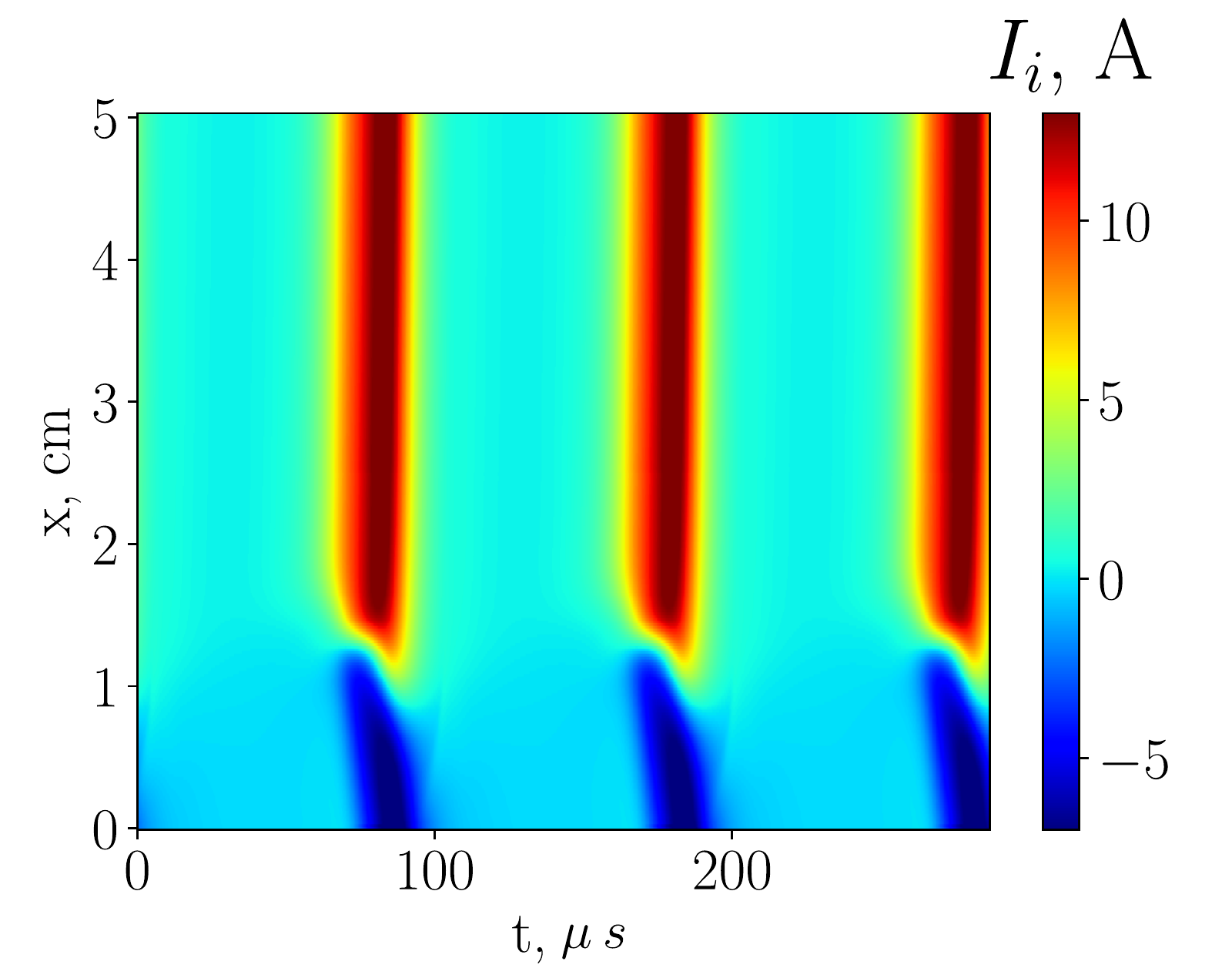}}
\subfloat[]{\includegraphics[width=0.5\textwidth]{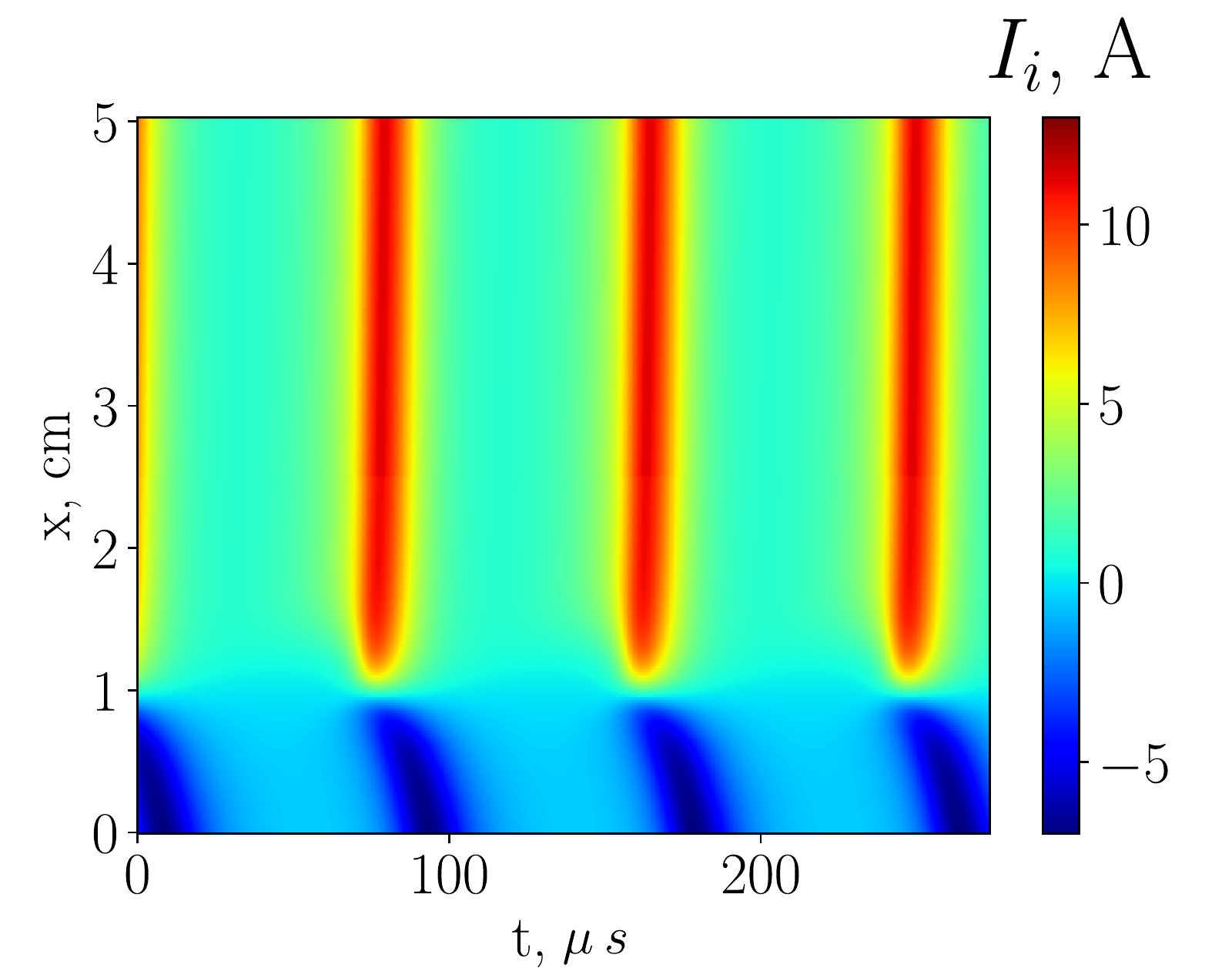}}
\caption{Temporal and spatial ion current dynamics in full model (a) and reduced model with lower $\beta$ (b), in Amperes (assuming cross section area $A = \SI{12.75\pi}{cm^2}$).}
\label{icur_cmpr}
\end{figure}

To show that the atom velocity $v_a$ plays a defining role in the observed oscillation frequency, the full and reduced model (lower $\beta$) were compared against each other, Fig.~\ref{va_effect}. Both full and reduced (with lower $\beta$) model show approximately linear behaviour for the frequency, again suggesting  the feedback loop mechanism supported by ion back-flow and advection of the neutral atoms. Interestingly, the ion current amplitudes are comparable between the full and reduced (lower $\beta$) models, Fig.~\ref{icur_cmpr}.

\section{Summary} \label{sec_summary}

In this paper, we have proposed a reduced continuum 1-D model with the ion velocity profile that has an ion back-flow (toward the anode) region. The model consists of two one-dimensional equations for coupled dynamics of the ion and neutral densities. While the 0-D simplification of this model,  commonly referred  to as the predator-prey model, shows the oscillations,  the continuum 1-D  model with a uniform  ion velocity and  boundary conditions typical of Hall thruster show damped modes converging to the analytical stationary state. Under externally imposed perturbations at the left boundary, the 1-D continuum model shows limited convective amplification of the perturbation (in the region where the local value of the neutral density is smaller than plasma density) and subsequent  convection (with a slow amplitude decay) where  the ion and neutral density perturbations are mostly decoupled.  Therefore the models with the uniform ion velocity, in the absence of additional effects e.g.\ such as temperature variations \cite{hara2014perturbation}, do not  predict the excitation of the breathing modes.     

The key element of the proposed  continuum model is  the ion velocity profile that has an ion back-flow (toward the anode) region. As it is shown here, such a system becomes unstable and shows self-consistent nonlinear oscillations. The comparison between this reduced model and the full self-consistent fluid model shows very similar characteristics for oscillations of the ion and neutral densities, and ion  current.  We have investigated the scaling of the frequency and amplitude of the oscillations as a function of the width of the back-flow region and neutral flow velocity  which show generic $v_a/L_b$ dependence both in the reduced and full models.
These results suggest that the presence of the ion back-flow region may be a  critical condition for the breathing mode oscillations and the overlap of this region with the ionization region creates the closed feedback loop necessary for the instability.  These results also indicate that the electron dynamics, in particular diffusion,  which leads to the appearance of the back-flow region, is important for the breathing mode. A number of full  models, reporting  the breathing mode oscillations, include the electron diffusion either explicitly via the electron fluid equations, such as full fluid \cite{barral2009low} or hybrid\cite{hagelaar2004modelling} formulations, or via full kinetic treatment as in the full kinetic (particle-in-cell) treatment \cite{coche2014two}. The ion back-flow near the anode occurs as a result of the electron diffusion ensuring the current conservation to maintain quasineutrality. The ion  stagnation in the back-flow region triggering enhanced ionization is suggested as the mechanism for the instability.   
It is also worth noting that the electron diffusion is also a reason for the appearance of the singular sonic point  where the ion velocity is equal to the local sound velocity, $v_i=c_s$. The smooth (regular) solution  is obtained when the sonic point is made regular by imposing some constraints on the current, mass flow rate, and plasma parameters. These constraints made singular point regular and play an important role for the existence and characteristics of stationary solutions \cite{dorf2003anode,CohenZurPoP2002,AhedoPoP2001,makowski2001review}. As it was shown in  \cite{smolyakov2019theory,RomadanovPPR2020}
the constraints indeed define the stability of the resulting profile and characteristic frequency of the breathing mode.

One has to note, however, that low frequency ionization oscillations were  also observed in the models without the electron diffusion, and hence, without back-flow region \cite{boeuf1998low,MorozovPPR2000a,MorozovPPR2000b,HaraPSST2018,hara2012one}. It has been suggested \cite{chable2005numerical}  that the high frequency oscillations due to the resistive axial modes \cite{chable2005numerical,FernandezPoP2008,KoshkarovPoP2018ax}  is  a driving mechanism  for a  breathing mode. Alternative mechanisms may be based on the temperature dependence of the ionization coefficient \cite{hara2014perturbation}, other electron temperature effects \cite{StaackAppPhysL2004}, and/or more complex interactions between the low frequency modes involving ionization and higher frequency modes in the ion transit time frequency range as it was suggested earlier  \cite{Kurzyna2008Plasma, smolyakov2019theory}. The simulations with the full model also demonstrate the existence of two different regimes of the breathing mode \cite{smolyakov2019theory}, the so-called solo regime, where only the low frequency mode present, and the regime with coexisting low frequency and the high frequency ion transit time oscillations. The breathing mode oscillations reproduced by our reduced model are similar to the solo regime, but are different 
from the modes coexisting with high-frequency ion transit time oscillations. The existence and nature of the breathing mode together with the well pronounced high-frequency mode is an interesting question that will be addressed in future studies with the full model. 

In summary, we would like to conclude that the importance of the reduced models is not only in their  ability to predict reasonably well some essential features of the ionization modes in Hall thrusters but also, and perhaps, even more importantly, in pointing  to some important  physics involved in the modes excitation and characteristics. The reduced models typically involve a smaller number of  adjustable parameters compared to the full models, e.g.\ as the values of the anomalous mobility are poorly known. Therefore, the tests of the  properties and consequences of the reduced models in the experiments \cite{Romadanov2018PSST,dale2019two1,dale2019two2,dale2019frequency} could be an effective approach to test the crucial physics responsible for the breathing modes oscillations. 

\acknowledgments{This work is supported in part by  US Air Force Office of
Scientific Research FA9550-15-1-0226, NSERC Canada, and Compute Canada computational resources.   The authors acknowledge with gratitude the important  discussions with I. V. Romadanov, J. B. Simmonds, I.D. Kaganovich, and K. Hara.}

\section*{Data availability}
The data that support the findings of this study are available from the corresponding author upon reasonable request.

\bibliography{ref}

\end{document}